%% file: jhep_draft_CG_SM.tex
\documentclass[a4paper,11pt]{article}
\usepackage{jheppub} 
\usepackage[T1]{fontenc} 
\RequirePackage[displaymath]{lineno} 

\usepackage{epsfig}
\usepackage{graphicx}
\usepackage{dcolumn}
\usepackage{bm}
\usepackage{ltablex,booktabs}
\usepackage{overpic}
\usepackage{subfigure}
\usepackage{float}
\usepackage{color}
\usepackage{amsmath}
\usepackage{mathcomp}
\usepackage{mathrsfs}
\usepackage{multirow}
\usepackage{rotating}
\usepackage{amssymb}
\usepackage{gensymb}
\usepackage{amsmath}
\usepackage{tabularx}
\usepackage{epstopdf}
\usepackage{makecell}
\usepackage{relsize}\usepackage{tabularx}
\usepackage{booktabs} 
\usepackage{xcolor} 
\def\babar{\mbox{\slshape B\kern-0.1em{\smaller A}\kern-0.1em B\kern-0.1em{\smaller A\kern-0.2em R}}}

\def\BF{\ensuremath{(1.026 \pm 0.008_{\rm{stat.}} \pm 0.009_{\rm{syst.}}) \%}}

\title{Amplitude analysis and branching fraction measurement of the decay \boldmath $D^0 \to K^0_S\pi^0\pi^0$}
\collaborationImg{\includegraphics[width=0.15\textwidth, angle=90]{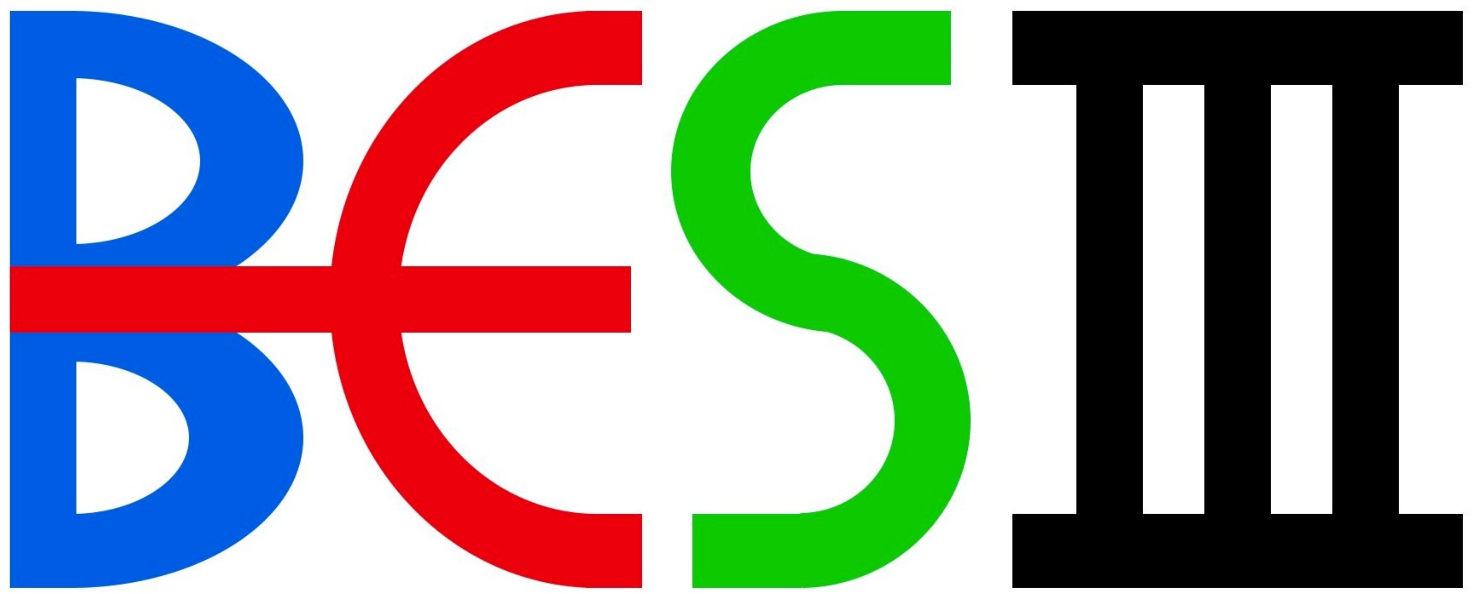}}
\collaboration{BESIII Collaboration}

\date{\today}
\abstract{
An amplitude analysis of the decay $D^0 \to K_S^0 \pi^0 \pi^0$ is performed to determine the relative magnitudes and phases of different intermediate processes. The analysis uses $e^+e^-$ collision data collected at the center-of-mass energy of 3.773~GeV by the BESIII detector corresponding to an 
integrated luminosity of 20.3~$\rm fb^{-1}$. 
The absolute branching fraction of $D^0 \to K^0_S \pi^0 \pi^0$ is measured to be \BF. The dominant intermediate process is $D^0 \to \bar{K}^{*}(892)^{0}(\to K^0_S \pi^0) \pi^0$, with a branching fraction of $(4.10\pm0.10_{\rm{stat.}}\pm0.07_{\rm{syst.}})\times 10^{-3}$.}

\arxivnumber{}
\begin{document}
\maketitle
\flushbottom

\section{Introduction}
Theoretical studies of hadronic decays in charm mesons are challenging due to the fact that the charm quark mass is neither heavy enough to support a reliable heavy quark mass expansion, nor light enough to allow for the application of chiral perturbation theory~\cite{Cheng}.
To address these challenges, non-perturbative methods are employed, which rely on precise experimental inputs to constrain model parameters, test theoretical predictions and guide the refinement of theoretical frameworks. This close interplay between theory and experiment drives the progress in understanding $D$-meson decays, particularly in studies of \emph{CP} violation~\cite{TDA_QCD}.

\begin{figure}[t]
\centering
    \subfigure[]{\includegraphics[width=0.48\textwidth]{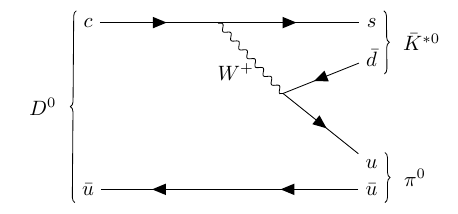}}
    \subfigure[]{\includegraphics[width=0.48\textwidth]{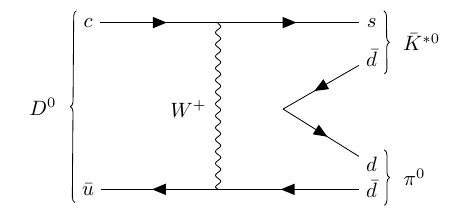}}
    \caption{Topological diagrams contributing to the decay $D^0 \rightarrow \bar K^{*}(892)^0 \pi^0$ with (a) color-suppressed internal $W$-emission tree diagram and (b) $W$-exchange diagram.}
\label{topology1}
\end{figure}

The lightest charmed mesons, known as the $D^{0}$ and $D^{+}$ mesons, decay exclusively through weak interactions. Their decay amplitudes are primarily governed by two-body processes such as $D \rightarrow VP$, $D \rightarrow PP$, $D \rightarrow SP$, and $D \rightarrow VV$, where $V$, $S$, and $P$ represent vector, scalar, and pseudoscalar mesons, respectively. In particular, the $D \rightarrow VP$ decays provide clearer opportunities compared to other two-body processes for elucidating the non-perturbative mechanism of charmed-meson decays~\cite{TDA_tree, TDA_QCD, Cheng}. In the decay $D^0 \to K_S^0\pi^0\pi^0$, a major contribution is expected to come from the Cabibbo-favored (CF) process $D^0 \to \bar K^{*}(892)^0\pi^0$, which is a typical $D \to VP$ decay. Within the topological-diagram approach (TDA)~\cite{TDA_tree}, CF decays provide critical inputs for determining topological amplitudes, which are subsequently used to predict singly Cabibbo-suppressed and doubly Cabibbo-suppressed modes and quantify SU(3) symmetry breaking. The corresponding topological diagrams of the $D^0 \to \bar K^{*}(892)^0\pi^0$ process, which are illustrated in Fig.~\ref{topology1}, can proceed via a color-suppressed internal $W$-emission tree diagram and a $W$-exchange diagram. Although the majority of theoretically predicted branching fractions (BFs) for the $D \to VP$ decays are in agreement with experimental measurements, there is an inconsistency observed in the BF for $D^0 \to \bar K^{*}(892)^0 \pi^0$. The predicted and measured values of the branching fractions are listed in Table~\ref{tab:theory}, where the predictions come from the pole model~\cite{pole}, the factorization-assisted topological-amplitude (FAT-mix) approach with $\rho-\omega$ mixing~\cite{FAT}, and the updated analysis of the two-body $D\rightarrow VP$ decays within the framework of the TDA~\cite{TDA}. 

Experimentally, the information on $D \to VP$ decays can be extracted through a three-body amplitude analysis. The CLEO collaboration has measured the BF of $D^0 \to \bar K^{*}(892)^0\pi^0$ to be $(2.74\pm0.23_{\rm{stat.}}\pm0.41_{\rm{syst.}})\%$ and $(4.16\pm0.37_{\rm{stat.}}\pm0.33_{\rm{syst.}})\%$ in the amplitude analysis of $D^0 \to K^- \pi^+ \pi^0$~\cite{kpipi0} and $D^0 \to K_S^0 \pi^0 \pi^0$~\cite{kspi0pi0}, respectively. These two experimental results show significant differences 
with each other and with the theoretical predictions. Hence, a more accurate
measurement of $D^0 \to \bar K^{*}(892)^0\pi^0$ is essential to offer a more rigorous test of the theoretical models and enhance the understanding of the dynamics in charmed-meson decays.

\begin{table}[htbp]
  \centering
  \begin{tabular}{|lc|}
    \hline
    Model  &$\mathcal{B}(D^0 \to \bar K^{*}(892)^0\pi^0)$~(\%)\\
    \hline	
    Pole~\cite{pole}     &2.9$\,\pm\,$1.0 \\
    FAT-mix~\cite{FAT} &3.25 \\
    TDA~\cite{TDA}      &3.61$\,\pm\,$0.18 \\    

    
    \hline
    CLEO~[from $D^0\to K^{-}\pi^+\pi^0$]~\cite{kpipi0} &$2.74\pm0.47$\\
    CLEO~[from $D^0\to K_{S}^{0}\pi^0\pi^0$]~\cite{kspi0pi0} &$4.16\pm0.49$\\
    \hline
  \end{tabular}
  \caption{The predicted and measured values of $\mathcal{B}(D^0 \to \bar K^{*}(892)^0\pi^0)$. The uncertainties of the CLEO measurements include both statistical and systematic contributions.}
  \label{tab:theory}
\end{table}

The amplitude analysis and BF measurement of the decay $D^0 \to K^0_S\pi^0\pi^0$ are presented in this paper, utilizing $e^+e^-$ collision data collected by the BESIII detector at BEPCII~\cite{lum1,lum2} corresponding to an integrated luminosity of 20.3 fb$^{-1}$. 
Charged-conjugate modes are always implied throughout this paper.
\section{Detector and data}
\label{sec:detector_dataset}
The BESIII  detector~\cite{Ablikim:2009aa} records symmetric $e^+e^-$ collisions provided by the BEPCII storage ring~\cite{Yu:IPAC2016-TUYA01} in the center-of-mass energy range from 1.84 to 4.95~GeV, with a peak luminosity of $1.1 \times 10^{33}\;\text{cm}^{-2}\text{s}^{-1}$ achieved at $\sqrt{s} = 3.773\;\text{GeV}$. BESIII has collected large data samples in this energy region~\cite{Ablikim:2019hff, EcmsMea, EventFilter}. The cylindrical core of the BESIII detector covers 93\% of the full solid angle and consists of a helium-based  multilayer drift chamber~(MDC), a plastic scintillator time-of-flight system~(TOF), and a CsI(Tl) electromagnetic calorimeter~(EMC), which are all enclosed in a superconducting solenoidal magnet providing a 1.0~T magnetic field. The solenoid is supported by an octagonal flux-return yoke with resistive plate counter muon identification modules interleaved with steel. The charged-particle momentum resolution at 1 GeV$/c$ is $0.5\%$, and the ionization energy loss (${\rm d}E/{\rm d}x$) resolution in the MDC is $6\%$ for electrons from Bhabha scattering. The EMC measures photon energies with a resolution of $2.5\%$ ($5\%$) at $1$~GeV in the barrel (end-cap) region. The time resolution in the TOF barrel region is 68~ps, while that in the end-cap region was 110~ps.  The end-cap TOF system was upgraded in 2015 using multigap resistive plate chamber technology, providing a time resolution of 60~ps, which benefits 86\% of the data used in this analysis~\cite{etof1,etof2,etof3}.

The data sample with a total integrated luminosity of $20.3 \, \mathrm{fb}^{-1}$, collected at the center-of-mass energy of $\sqrt{s} = 3.773\, \mathrm{GeV}$, is used in this analysis. The $\psi(3770)$ predominantly decays to $D^+D^-$ or $\bar{D}^0 D^0$ pairs without additional hadronic activity, providing an ideal environment for studying $D$ meson decays with the double-tag (DT) technique~\cite{DTmethod}.
In this method, a single-tag (ST) candidate requires the reconstruction of a single $\bar D^0$ meson through one of the three hadronic decay modes: $\bar D^0 \to K^+\pi^-$, $\bar D^0 \to K^+\pi^-\pi^0$ and $\bar D^0 \to K^+\pi^-\pi^-\pi^+$. In a DT candidate, both the $D^0$ and $\bar D^0$ mesons are reconstructed, with the $D^0$ meson decaying to the signal mode $D^0 \to K_S^0\pi^0\pi^0$ and the $\bar D^0$ meson decaying to one of the ST modes, which is referred to as the tag mode.

Monte Carlo (MC) simulated data samples, produced with a {\sc
geant4}-based~\cite{geant4} software package, 
are used to determine detection efficiencies and estimate backgrounds. 
The simulation includes the geometric description of the BESIII detector, the
detector response, and models the beam-energy spread and initial-state radiation (ISR) in the $e^+e^-$
annihilations with the generator {\sc kkmc}~\cite{KKMC1,KKMC2}. The inclusive MC sample includes the production of $D\bar{D}$ pairs (with quantum-coherence effects for the neutral $D^0 \bar D^0$ channels), the non-$D\bar{D}$ decays of the $\psi(3770)$, the ISR production of the $J/\psi$ and $\psi(3686)$ states, and the continuum processes incorporated in {\sc kkmc}~\cite{KKMC1,KKMC2}.
All particle decays are modeled with {\sc
evtgen}~\cite{EVTGEN1,EVTGEN2} using BFs 
either taken from the
Particle Data Group (PDG)~\cite{PDG}, when available,
or otherwise estimated with {\sc lundcharm}~\cite{LUNDCHARM1,LUNDCHARM2}.
Final-state radiation
from charged final-state particles is incorporated using {\sc
photos}~\cite{PHOTOS}.
A phase-space (PHSP) MC sample is generated with a uniform distribution for the decay $D^0 \to K_S^0 \pi^0 \pi^0$, which is used to determine the detection efficiency function mentioned in Sec.~\ref{sec:fitmethod} and calculate the normalization integral used in the determination of the amplitude-model parameters in the fit to data.
A signal MC sample, generated according to the results of the amplitude analysis for the decay $D^0 \to K_S^0 \pi^0 \pi^0$, is used to check the fit performance, calculate the goodness of fit and estimate the average DT efficiency in the BF measurement.

\section{Event selection}
\label{ST-selection}

The $D^0$ candidates are constructed from individual $\pi^\pm, \pi^0, K^\pm$ and $K_S^0$ candidates with the following selection criteria, which are the common requirements for both the amplitude analysis and BF measurement. Additional requirements used in the amplitude analysis are discussed in Sec. \ref{sec:pwa-select}. 
 
Charged tracks detected in the MDC are required to be within a polar angle ($\theta$) range of $|\rm{cos\theta}|<0.93$, where $\theta$ is defined with respect to the $z$-axis, which is the symmetry axis of the MDC. For charged tracks, the distance of closest approach to the interaction point (IP) must be less than 10\,cm along the $z$-axis,  and less than 1\,cm in the transverse plane.

Particle identification~(PID) for charged tracks combines measurements of the ionization energy loss d$E$/d$x$ and the flight time in the TOF to form likelihoods $\mathcal{L}(h)~(h=K,\pi)$ for each hadron $h$ hypothesis.
Charged kaons and pions are identified by comparing the likelihoods, $\mathcal{L}(K)>\mathcal{L}(\pi)$ and $\mathcal{L}(\pi)>\mathcal{L}(K)$, respectively.

The $K_{S}^0$ candidates are selected from all possible pairs of tracks with opposite charges and the distances of the charged tracks to the interaction point along the beam direction are required to be within 20 cm. The selected charged tracks are assigned as pions and no further PID requirements are applied.
A primary vertex and a secondary vertex are reconstructed, and the decay length between two vertices is required to be greater than twice its uncertainty. The $\chi^2$ of the vertex fit must be less than 100 and the invariant mass of the $\pi^{+}\pi^{-}$ pair ($M_{\pi^{+}\pi^{-}}$) is required to be in the range $[0.487, 0.511]$ GeV/$c^2$ to form the candidate $K_{S}^0$ particles.

Photon candidates are identified using isolated showers in the EMC.  The deposited energy of each shower must be more than 25~MeV in the barrel region ($|\cos\!\theta|< 0.80$) and more than 50~MeV in the end-cap region ($0.86 <|\cos\!\theta|< 0.92$).  To exclude showers that originate from charged tracks, the angle between the EMC shower and the position of the closest charged track at the EMC must be greater than 10 degrees as measured from the IP. To suppress electronic noise and showers unrelated to the event, the difference between the EMC time and the event start time is required to be within [0, 700]\,ns.

The $\pi^0$ candidates are formed from the photon pairs with invariant masses in a range of $[0.115, 0.150]$~GeV/$c^2$, which is about three times the mass resolution. Moreover, in order to achieve an adequate resolution, at least one of the two photons is required to be detected in the barrel EMC. A kinematic fit that constrains the $\gamma\gamma$ invariant mass to the known $\pi^{0}$ mass~\cite{PDG} is performed to improve the mass resolution. The $\chi^2$ of the kinematic fit is required to be less than 50.

To distinguish the $D^0 \bar D^0$ mesons from the backgrounds, the beam-constrained mass~($M_{\rm{BC}}$) and the energy difference~($\Delta E$) are employed. They are defined as
\begin{equation}
\begin{aligned}
	&M_{\rm{BC}} = \sqrt{E_{\rm{beam}}^2 /c^4-|\vec{p}_{D}|^2 /c^2}, \\
	&\Delta E = E_{D} - E_{\rm{beam}},
\end{aligned}
\label{MBC_DeltaE}
\end{equation}
where $\vec{p}_{D}$ and $E_{D}$ represent the total reconstructed momentum and energy of the $D$ candidate, and $E_{\rm{beam}}$ is the beam energy. The $D$ signal manifests itself as a peak around the known $D$ mass~\cite{PDG} in the $M_{\rm{BC}}$ distribution and as a peak around zero in the $\Delta E$ distribution. For each tag mode, if there are multiple combinations, the one giving the minimum $|\Delta E_{\text{tag}}|$ is retained for further analysis. The signal $D^0$ candidates are reconstructed from the particles that have not been used for the tagged $\bar D^0$ reconstruction, with $K_S^0$ and $\pi^0$ reconstructed through $\pi^+ \pi^-$ and $\gamma \gamma$, respectively. They are identified using the energy difference and the beam-constrained mass of the signal side, $|\Delta E_{\text{sig}}|$ and $M^{\text{sig}}_{\mathrm{BC}}$. If there are multiple combinations, the one giving the minimum $|\Delta E_{\text{sig}}|$ is retained for further analysis. 

In order to enhance the selection efficiencies of $D$ mesons and effectively suppress background candidates, mode-dependent requirements on the energy difference $\Delta E$ are applied to both the signal and tag modes. The corresponding $\Delta{E}$ regions for the signal and each tag mode are provided in Table~\ref{tab:deltaEcut}.

\begin{table}[hbtp]
  \begin{center}
     \centering \begin{tabular}{|lc|}
      \hline
      Decay mode &$\Delta{E}~(\rm {GeV})$\\
      \hline
      $D^0\to K_{S}^{0}\pi^0\pi^0$ &(-0.070, 0.030)\\
      \hline
      $\bar {D}^0\to K^+\pi^-$  &(-0.027, 0.027)\\
      $\bar {D}^0\to K^+\pi^-\pi^0$  &(-0.062, 0.049)\\
      $\bar {D}^0\to K^+\pi^-\pi^-\pi^+$  &(-0.026, 0.024)\\
      \hline
    \end{tabular}
    \caption{Requirements of $\Delta{E}$ for the signal and the different $\bar D^0$ tag modes.}
    \label{tab:deltaEcut}
  \end{center}
\end{table}

\section{Amplitude analysis}
\label{Amplitude-Analysis}
\subsection{Additional selection criteria in the amplitude analysis}
\label{sec:pwa-select}

To enhance the signal purity for the amplitude analysis, candidate events are selected with $1.858 < M_{\rm BC} < 1.872~{\rm GeV}/c^2$ for the signal side and $1.859 < M_{\rm BC} < 1.872~{\rm GeV}/c^2$ for the tag modes. The sources of background for the $D^0 \to K^0_S \pi^0 \pi^0$ candidates are investigated by analyzing the inclusive MC sample. The backgrounds are identified and categorized into two sources: the peaking backgrounds and the misidentified backgrounds. The peaking backgrounds, originating from $D^0 \to K^0_S K^0_S$  (where one $K_S^0$ meson decays via $K_S^0 \to \pi^0 \pi^0$) and $D^0 \to \pi^+ \pi^- \pi^0 \pi^0$, are modeled using the inclusive MC sample in the two-dimensional (2D) fit to the $M_{\rm{BC}}^{\rm{sig}}$ versus $M_{\rm{BC}}^{\rm{tag}}$ distribution (see Appendix~\ref{2dfit} for details). 
The first type of misidentified background arises from the reconstruction of $\psi(3770) \to D^+ D^-$ events as $D^0 \bar{D}^0$ pairs. This occurs when a $\pi^+$ meson from the decay $D^+ \to K_S^0\pi^0\pi^+$ is swapped with a $\pi^0$ meson from the decay $D^- \to K^+\pi^-\pi^-\pi^0$. As a result, the $D^+$ is falsely identified as a $D^0 \to K_S^0\pi^0\pi^0$, while the $D^-$ is misidentified as a $\bar{D}^0 \to K^+\pi^-\pi^-\pi^+$. To suppress this background, events simultaneously satisfying the conditions $1.839<M_{K_S^0\pi^+\pi^0_1}<1.899$ GeV/$c^2$ and $1.839<M_{K^+\pi^-\pi^-\pi^0_2}<1.899$ GeV/$c^2$ are rejected, where the momentum of $\pi^0_1$ is higher than that of $\pi^0_2$. 
The second type of misidentified background arises from the reconstruction of $\psi(3770) \to D^0 \bar D^0$ events as other decay modes of $D^0 \bar{D}^0$ pairs. This occurs when a $\pi^+ \pi^- $ pair from the decay $D^0 \to K_S^0 \pi^+ \pi^- \pi^0$ is swapped with a $\pi^0$ meson from the decay $\bar D^0 \to K^+\pi^-\pi^0$. As a result, the $D^0$ is falsely identified as a $D^0 \to K_S^0\pi^0\pi^0$, while the $\bar D^0$ is misidentified as a $\bar{D}^0 \to K^+\pi^-\pi^-\pi^+$. To suppress this background, events simultaneously satisfying the conditions $1.835<M_{K_S^0\pi^-\pi^+\pi^0}<1.895$ GeV/$c^2$ and $1.835<M_{K^+\pi^-\pi^0}<1.895$ GeV/$c^2$ are rejected.

To achieve optimal resolution and ensure that all events are within the PHSP boundary, a kinematic fit is performed, in which the four-momenta of the final-state particles are constrained to the initial four-momenta of the $e^+e^-$ system and the reconstructed masses of the $D^0$, $K_S^0$ and $\pi^0$ mesons are constrained to their known values~\cite{PDG}. The four-momenta of the final-state particles are updated by the kinematic fit for the amplitude analysis. 

\begin{figure}[htbp]
  \centering
 \includegraphics[width=0.90\textwidth]{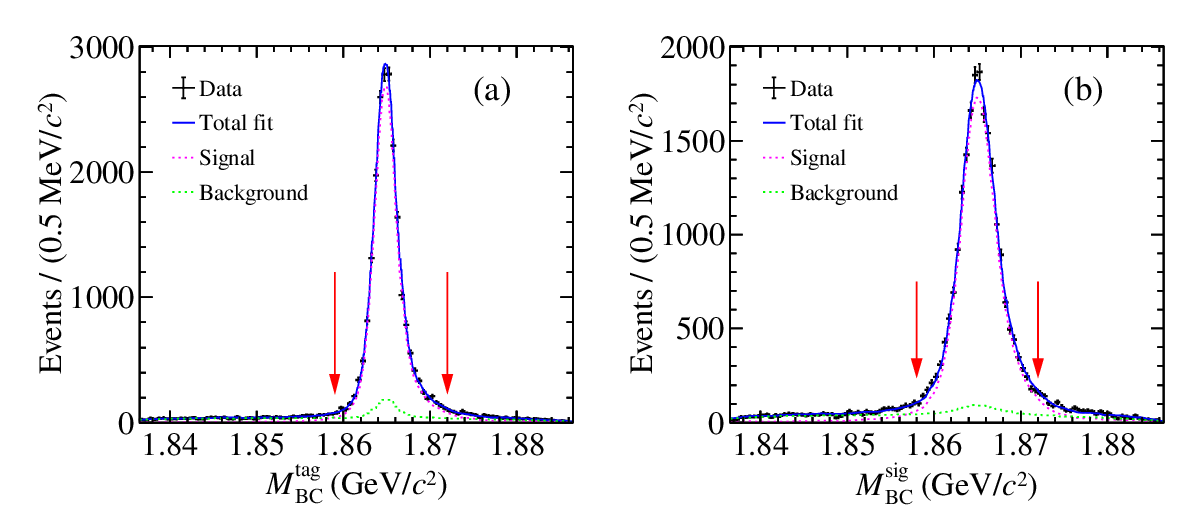}
  \caption{Projections on $M_{\rm BC}^{\rm tag}$~(a) and $M_{\rm BC}^{\rm sig}$~(b) of the 2D fit. The points with error bars are data. The solid blue line is the total fit result, the dotted red line is the signal, the dotted green line is the background and the red arrows represent the signal region.}
  \label{fig:2dfit}
\end{figure}

After applying all of the aforementioned criteria, there are 20349 events retained in the signal region for the amplitude analysis. The signal purity ($\omega_{\rm{sig}}$), determined from an unbinned 2D maximum likelihood fit to the $M_{\rm{BC}}^{\rm{sig}}$ versus $M_{\rm{BC}}^{\rm{tag}}$ distribution (see Appendix~\ref{2dfit} for details), is measured to be $(93.5\pm 0.2)\%$. The purity uncertainty is obtained by propagating uncertainties of fitted parameters according to the correlation matrix. The fit results are shown in Fig. \ref{fig:2dfit}.

\subsection{Fit method}
\label{sec:fitmethod}

An unbinned maximum-likelihood fit is used in the amplitude analysis of the $D^0 \to K^0_S\pi^0\pi^0$ decay. The likelihood function $\mathcal{L}$ is constructed  by incoherently adding the background probability density function (PDF) to the signal PDF. After taking the logarithm, the combined PDF can be written as

\begin{eqnarray}\begin{aligned}
\ln{\mathcal{L}} = \begin{matrix}\sum\limits_{k=1}^{N_{\rm data}} \ln [\omega_{\rm sig}f_{S}(p^{k})+(1-\omega_{\rm sig})f_B(p^k)]\end{matrix}\,,
\label{likelihood3}
\end{aligned}\end{eqnarray}
where $k$ indicates the $k^{\rm th}$ event in the data sample, $N_{\rm data}$ is the number of retained events, $p$ denotes the four-momenta of the final state particles, $f_{S} (f_{B})$ is the signal (background) PDF and $\omega_{\rm sig}$ is the signal purity discussed in Sec.~\ref{sec:pwa-select}.

The signal PDF $f_S$ is given by 
\begin{eqnarray}\begin{aligned}
  f_S(p) = \frac{\epsilon(p)\left|\mathcal{M}(p)\right|^{2}R_{3}(p)}{\int \epsilon(p)\left|\mathcal{M}(p)\right|^{2}R_{3}(p)\,{\rm d}p}\,, \label{signal-PDF}
\end{aligned}\end{eqnarray}
where $\epsilon(p)$ is the detection efficiency parameterized in terms of the four-momenta $p$, and $R_{3}(p)$ is the PHSP factor for three-body decays. This is defined as
\begin{equation}
R_3(p) = \delta\left(p_{D^0} - \sum_{j=1}^3 p_j\right) \prod_{j=1}^3 \delta(p_j^2 - m_j^2) \theta(E_j),
\end{equation}
where $j$ runs over the three daughter particles, $E_j$ is the energy of particle $j$ and $\theta(E_j)$ is the step function. The total amplitude $\mathcal{M}$ is modeled with the isobar model, which is a coherent sum of the individual amplitudes of intermediate processes and is given by $\mathcal{M}=\sum \rho_{n}e^{i\phi_{n}}\mathcal{A}_{n}$. The magnitude $\rho_{n}$ and phase $\phi_{n}$ are the free parameters to be determined by the fit. The amplitude of the $n^{\rm th}$ intermediate process ($\mathcal{A}_{n}(p)$) is given by

\begin{eqnarray}
\begin{aligned}
  \mathcal A_n(p_j) = P_n S_n X_n^r X_n^D\,, \label{base-amplitude}
\end{aligned}\end{eqnarray}
where $X_{n}^{r}$ and $X_{n}^{D}$ are the Blatt-Weisskopf barrier for the intermediate resonances and the $D^0$ meson, respectively (Sec.~\ref{Blatte}), $P_{n}$ is the propagator of the intermediate resonance (Sec.~\ref{sec:propagator}), and $S_{n}$ is the spin factor constructed with the covariant tensor formalism~\cite{covariant-tensors} (Sec.~\ref{sec:spinfactor}).

The background PDF is given by 
\begin{equation}
  f_B(p)=\frac{B(p)R_3(p)}{\int{B(p)R_3(p)}dp},	
  \label{pdf:bkg}
\end{equation}
where $B(p)$ represents the background function. In the numerator of Eq.~(\ref{signal-PDF}), the $\epsilon(p)$ and $R_3(p)$ terms,  which are independent of the fitted variables, are treated as constants and can be neglected. To extract the shared component $\epsilon(p)R_3(p)$, the background PDF can be expressed as
\begin{equation}
  f_B(p)=\frac{\epsilon(p)B_{\epsilon}(p)R_3(p)}{\int{\epsilon(p)B_{\epsilon}(p)R_3(p)}dp},	
  \label{pdf:bkg_new}
\end{equation}
where $B_{\epsilon}(p) = B(p) / \epsilon(p)$  is the efficiency-corrected background function. The background function in the data is modeled by the background events in the signal region derived from the inclusive MC sample. The invariant-mass distributions of events outside the signal region show good agreement between the data and MC simulation, thus validating the description from the inclusive MC sample. The distributions of background events from the inclusive MC sample have been examined both inside and outside the signal region. Generally, they are compatible with each other within statistical uncertainties.

The invariant-mass squared distributions $(M^2_{K^0_S\pi^0_1}, M^2_{K^0_S\pi^0_2}, M^2_{\pi_1 ^0 \pi_2 ^0})$ of background events in inclusive MC sample are modeled using the XGBoost package~\cite{XG1,XG2}. The candidates in the signal region are used as a training sample, and two different models are trained. The first XGBoost model is utilized to predict the background probability $B(p)R_3(p)$, where the training samples are the background events from the inclusive MC sample. Subsequently, a separate XGBoost model is employed to predict the efficiency probability $\epsilon(p)R_3(p)$, with the training dataset derived from PHSP MC samples that have undergone reconstruction and event selection procedures. Dividing the two probabilities results in the value of $B_{\epsilon}(p)$, which is equal to $B(p)/\epsilon(p)$.

Through combining Eq. (\ref{signal-PDF}) with Eq. (\ref{pdf:bkg_new}) and neglecting the term $\epsilon(p)R_3(p)$, the log-likelihood becomes
\begin{equation}
\ln\mathcal{L}=\sum\limits_{k}^{N_{\rm {data}}}\ln\bigg[\omega_{\rm {sig}}\frac{|\mathcal{M}(p)|^2}{\int \epsilon(p)|\mathcal{M}(p)|^2R_3(p)dp}+(1-\omega_{\rm {sig}})\frac{B_{\epsilon}(p)}{\int{\epsilon(p)B_{\epsilon}(p)R_3(p)}dp}\bigg].
  \label{likelihoodexa}
\end{equation}

The normalization integrals of signal and background are evaluated with signal MC samples,
\begin{eqnarray}\begin{aligned}
  \int \epsilon(p) |\mathcal{M}(p)|^2 R_{3}(p)\,{\rm d}p \approx
\frac{1}{N_{\rm MC}}\sum_{k_{\rm MC}}^{N_{\rm MC}} \frac{ |\mathcal{M}(p^{k_{\rm MC}})|^2 }{\left|\mathcal{M}^{g}(p^{k_{\rm MC}})\right|^{2}}\,, \label{MC-intergral}
\end{aligned}\end{eqnarray}

\begin{eqnarray}\begin{aligned}
  \int \epsilon(p) B_{\epsilon}(p) R_{3}(p)\,{\rm d}p \approx
\frac{1}{N_{\rm MC}}\sum_{k_{\rm MC}}^{N_{\rm MC}} \frac{ B_{\epsilon}(p^{k_{\rm MC}}) }{\left|\mathcal{M}^{g}(p^{k_{\rm MC}})\right|^{2}}\,, \label{bkgintergral}
\end{aligned}\end{eqnarray}
where $k_{\rm MC}$ is the index of the $k^{\rm th}$ event of the signal MC sample, and $N_{\rm MC}$ is the number of the selected signal MC events. 
The symbol $\mathcal{M}^{g}(p)$ denotes the PDF used to generate the signal MC sample in the MC integration. Here, the signal is the $D^0 \to K_S^0 \pi^0 \pi^0$ decay that have passed reconstruction and event selection procedures. The computational efficiency of the MC integration is significantly improved by evaluating the normalization integral with signal MC samples. These samples intrinsically take into account the event selection acceptance and the detection resolution.

To account for the bias caused by differences in $K_S^0$ and $\pi^0$ reconstruction between data and MC simulation, each signal MC event is weighted with a ratio, $\gamma_{\epsilon}(p)$, which is calculated as
\begin{equation}
  \gamma_{\epsilon}(p) = \frac{{\epsilon_{K_S^0,\rm data}(p)}{\epsilon_{\pi^0,\rm data}(p)}{\epsilon_{\pi^0,\rm data}(p)}}{{\epsilon_{K_S^0,\rm MC}(p)}{\epsilon_{\pi^0,\rm MC}(p)}{\epsilon_{\pi^0,\rm MC}(p)}},
  \label{pwa:gamma}
\end{equation}
where $\epsilon_{K_S^0,\rm data}(p)$ and $\epsilon_{\pi^0,\rm MC}(p)$ are the $K^0_S$ and $\pi^0$ efficiencies as a function of the momenta of the daughter particles for data and MC simulation, respectively. Then the MC integration is determined by 
\begin{eqnarray}\begin{aligned}
    &\int \epsilon(p) |\mathcal{M}(p)|^2 R_{3}\,{\rm d}p \approx
&\frac{1}{N_{\rm MC}} \sum_{k_{\rm MC}}^{N_{\rm MC}} \frac{ |\mathcal{M}(p^{k_{\rm MC}})|^2 \gamma_{\epsilon}(p^{k_{\rm MC}})}{\left|\mathcal{M}^{g}(p^{k_{\rm MC}})\right|^{2}}\,,
\label{MC-intergral-corrected}
\end{aligned}\end{eqnarray}

\begin{eqnarray}\begin{aligned}
    &\int \epsilon(p) B_{\epsilon}(p) R_{3}\,{\rm d}p \approx
&\frac{1}{N_{\rm MC}} \sum_{k_{\rm MC}}^{N_{\rm MC}} \frac{ B_{\epsilon}(p^{k_{\rm MC}}) \gamma_{\epsilon}(p^{k_{\rm MC}})}{\left|\mathcal{M}^{g}(p^{k_{\rm MC}})\right|^{2}}\,.
\label{bkg-intergral-corrected}
\end{aligned}\end{eqnarray}

\subsubsection{Blatt-Weisskopf barrier factors}\label{Blatte}
The Blatt-Weisskopf barrier factors $X_L(q)$~\cite{Blatte} are the barrier functions for a two-body decay process $a \to bc$. These functions depend on the angular momentum $L$ and the momenta $q$ of the final-state particle $b$ or $c$ in the rest system of $a$. They are taken as
\begin{eqnarray}
\begin{aligned}
 X_{L=0}(q)&=1,\\
 X_{L=1}(q)&=\sqrt{\frac{z_0^2+1}{z^2+1}},\\
 X_{L=2}(q)&=\sqrt{\frac{z_0^4+3z_0^2+9}{z^4+3z^2+9}}\,,
\end{aligned}
\end{eqnarray}
where $z=qR_r$, $z_0=q_0R_r$ and the effective radius, $R_r$, of the barrier is fixed to 3.0~$({\rm GeV}/c)^{-1}$ for the intermediate 
resonances and 5.0~$({\rm GeV}/c)^{-1}$ for the $D^0$ meson. The momentum $q$ is given by
\begin{eqnarray}
\begin{aligned}
q = \sqrt{\frac{(s_a+s_b-s_c)^2}{4s_a}-s_b}\,, \label{q2}
\end{aligned}
\end{eqnarray}
where $s_a, s_b, \text{and}~s_c$ are the invariant-mass squared of particles $a, b, \text{and}~c$, respectively. The value of $q_0$ is that of $q$ when $s_a = m_a^2$, where $m_a$ is the mass of particle $a$.


\subsubsection{Propagator}\label{sec:propagator}
The intermediate resonances $\bar K^{*}(892)^{0}$, $\bar K^{*}(1410)^{0}$, $\bar K_2^{*}(1430)^{0}$, $\bar K^{*}(1680)^{0}$ and $f_2(1270)$ are
parameterized with a relativistic Breit-Wigner function,
\begin{eqnarray}\begin{aligned}
    P(m) = \frac{1}{m_{0}^{2} - m^2 - im_{0}\Gamma(m)/c^2}\,,\; 
    \Gamma(m) = \Gamma_{0}\left(\frac{q}{q_{0}}\right)^{2L+1}\left(\frac{m_{0}}{m}\right)X^{2}_{L}(q)\,, 
  \label{RBW}
\end{aligned}\end{eqnarray}
where $m$ is the invariant mass of the decay products, $m_0$ and $\Gamma_0$ are the mass and width of the intermediate resonance that are fixed to their known values~\cite{PDG}. The energy-dependent width is denoted by $\Gamma(m)$. The $X_L(q)$ is the Blatt-Weisskopf barrier factor, defined in Sec.~\ref{Blatte}.

The $(K_S^0\pi^0)_{S\rm-wave}$ is modeled using the K-matrix parameterization with the same formula as in Ref.~\cite{Kpisw}. The amplitude has two isospin parts $\mathcal{A}_{1/2}$ and $\mathcal{A}_{3/2}$. The form for $I = 1/2$ is 
\begin{equation}
\mathcal{A}_{1/2} = \alpha_{K\pi} \hat{T}_{11} + \alpha_{K\eta^\prime} \hat{T}_{12}, 
\end{equation}
where
\begin{align}
\hat{T} &= (I - i\hat{K}\rho)^{-1}\hat{K} \\
&= \frac{1}{1 - \rho_1\rho_2\hat{D}-i(\rho_1K_{11}+\rho_2K_{22})}
\begin{pmatrix}
K_{11}-i\rho_2\hat{D} & K_{12} \\
K_{21} & K_{22}-i\rho_1\hat{D}
\end{pmatrix}
\end{align}
and $\hat{D} = \det(\hat{K}) = K_{11}K_{22}-K_{12}^2$. $\hat{K}$ represents a $2\times2$ matrix~\cite{FOCUS}, which includes two channels of $K\pi$ and $K\eta^\prime$. The elements of $\hat{K}$ are
\begin{align}
K_{11} &= \left(\frac{s - s_{0\frac{1}{2}}}{s_{\text{norm}}}\right)\left(\frac{g_1\cdot g_1}{s_1 - s}+C_{110}+C_{111}\tilde{s}+C_{112}\tilde{s}^2\right), \\
K_{22} &= \left(\frac{s - s_{0\frac{1}{2}}}{s_{\text{norm}}}\right)\left(\frac{g_2\cdot g_2}{s_1 - s}+C_{220}+C_{221}\tilde{s}+C_{222}\tilde{s}^2\right), \\
K_{12} &= \left(\frac{s - s_{0\frac{1}{2}}}{s_{\text{norm}}}\right)\left(\frac{g_1\cdot g_2}{s_1 - s}+C_{120}+C_{121}\tilde{s}+C_{122}\tilde{s}^2\right),
\end{align}
where $s = m_{K_S^0\pi^0}^2$ and the factor of $s_{\text{norm}} = m_K^2 + m_\pi^2$ is introduced to make the expression dimensionless. $g_1$ and $g_2$ are the coupling constants and $\tilde{s} = s/s_{\text{norm}}$. The rest parameters are listed in Table~\ref{tab:K-Matrix}. $\alpha_{K\pi}$ and $\alpha_{K\eta^\prime}$ are free parameters. The form for $I = 3/2$ is
\begin{equation}
\mathcal{A}_{3/2} = F_{3/2} = \frac{1}{1 - iK_{3/2}\rho_{K\pi}},
\end{equation}
where
\begin{equation}
K_{3/2} = \left(\frac{s - s_{0\frac{3}{2}}}{s_{\text{norm}}}\right)\left(D_{110}+D_{111}\tilde{s}+D_{112}\tilde{s}^2\right)
\end{equation}
and the phase space factor $\rho$ is
\begin{equation}
\rho_{K\pi}(s) = \sqrt{\left(1 - \frac{(m_K + m_\pi)^2}{s}\right)\left(1 - \frac{(m_K - m_\pi)^2}{s}\right)}.
\end{equation}

\begin{table}[htbp]
\begin{center}
\begin{tabular}{|lllll|}
\hline
Pole($\rm GeV^2$) & Coupling($\rm GeV$) & $C_{11i}$ & $C_{12i}$ & $C_{22i}$ \\
\hline
$s_1 = 1.7919$ &  &  & & \\
& $g_1 = 0.31072$ &  &  &  \\
 & $g_2 = -0.02323$ &  &  &  \\
 &  & $C_{110} = 0.79299$ & $C_{120} = 0.15040$ & $C_{220} = 0.17054$ \\
 &  & $C_{111} = -0.15099$ & $C_{121} = -0.038266$ & $C_{221} = -0.0219$ \\
 &  & $C_{112} = 0.00811$ & $C_{122} = 0.0022596$ & $C_{222} = 0.00085655$ \\
\hline
\end{tabular}
\caption{Values of constant parameters for the $\rm I = 1/2$ K-matrix~\cite{FOCUS}.}
\label{tab:K-Matrix}
\end{center}
\end{table}

As a result, $(K_S^0\pi^0)_{S\rm-wave}$ can be expressed as $\mathcal{A}_{1/2}+\alpha_{3/2}\mathcal{A}_{3/2}$. The complex coefficients $\alpha_{K\pi}$, $\alpha_{K\eta^{\prime}}$ and $\alpha_{32}$ are the free parameters in the fit.

The $\pi^0\pi^0~S$-wave is modeled by K-matrix parametrization. Detailed descriptions of the K-matrix formalism can be found in various references~\cite{km1,km2,km3,km4}. The term ``K-matrix amplitude'' refers to the product of the production vector $P$ and the matrix propagator $(I-iK\rho)^{-1}$:
\begin{equation}
    A_i=(I-iK\rho)^{-1}_{ij}P_j,
  \end{equation}
where $I$ is identity matrix, $K$ is the K-matrix describing the scattering process and $\rho$ is the PHSP matrix. The indices $i$ and $j$ represent the coupled channels ($1 = \pi\pi$, $2 = K\Bar{K}$, $3 = 4\pi$, $4 = \eta\eta$, $5 = \eta\eta^{\prime}$).

The K-matrix is expressed as
\begin{equation}
    K_{ij}(s)=\bigg (\sum_{\alpha}\frac{g_i^{\alpha}g_j^{\alpha}}{m_{\alpha}^2-s}+f_{ij}^{\rm scatt}\frac{1~{\rm GeV}^{2}/c^4 -s_0^{\rm scatt}}{s-s_0^{\rm scatt}}\bigg)\Bigg[\frac{1~{\rm GeV}^{2}/c^4 -s_{A_0}}{s-s_{A_0}}\bigg(s-s_{A}m_{\pi}^{2}/2\bigg)\Bigg],
  \end{equation}
where $s=m^{2}_{\pi^0 \pi^0}$ and $g_i^{\alpha}$ denote the real coupling constants of the pole $m_{\alpha}$ to meson channel $i$. The parameters $f_{ij}^{\rm scatt}$ and $s_0^{\rm scatt}$ describe a smooth part for the K-matrix elements. 

The $P$ vector is given by 
\begin{equation}
    P_{j}(s)=f_{1j}^{\rm prod}\frac{1~{\rm GeV}^{2}/c^4 -s_0^{\rm prod}}{s-s_0^{\rm prod}}+\sum_{\alpha}\frac{\beta^{\alpha}g_j^{\alpha}}{m_{\alpha}^2-s},
  \end{equation}
where $f_{1j}^{\rm prod}$ and $\beta_{\alpha}$ describe the production of the slowly varying part of the K-matrix. In this analysis, the K-matrix can be completely fixed using the values of all these parameters from the amplitude analysis of $D^0\to K_S^0\pi^+\pi^-$ in the \babar~and Belle experiments~\cite{KPsnew2}, except for $\beta_1$ and $f_{11}^{\mathrm{prod}}$.

\subsubsection{Spin factors}\label{sec:spinfactor}
Due to the limited size of the PHSP, the analysis is restricted to intermediate resonances with spin $J = 0, 1$, and $2$. In the decay process $a \to bc$, exclusive consideration is given to systems with orbital angular momentum $L = 0, 1$, and $2$, as these configurations dominate the observed decay modes. The momenta of the particles $a$, $b$, and $c$ in the $a \to bc$ process are denoted by $p_a$, $p_b$, and $p_c$, respectively.

The spin-projection operators~\cite{covariant-tensors} are defined as
\begin{eqnarray}
\begin{aligned}
  &P^{(0)}(a) = 1\,,&(S~\rm{wave})\\
  &P^{(1)}_{\mu\mu^{\prime}}(a) = -g_{\mu\mu^{\prime}}+\frac{p_{a,\mu}p_{a,\mu^{\prime}}}{p_{a}^{2}}\,,&(P~\rm{wave})\\
  &P^{(2)}_{\mu\nu\mu^{\prime}\nu^{\prime}}(a) = \frac{1}{2}(P^{(1)}_{\mu\mu^{\prime}}(a)P^{(1)}_{\nu\nu^{\prime}}(a)+P^{(1)}_{\mu\nu^{\prime}}(a)P^{(1)}_{\nu\mu^{\prime}}(a)) -\frac{1}{3}P^{(1)}_{\mu\nu}(a)P^{(1)}_{\mu^{\prime}\nu^{\prime}}(a)\,.&(D~\rm{wave})
 \label{spin-projection-operators}
\end{aligned}
\end{eqnarray}
The pure orbital angular-momentum covariant tensors are given by 
\begin{eqnarray}
\begin{aligned}
    \tilde{t}^{(0)}_{\mu}(a) &= 1\,,&(S~\rm{wave})\\
    \tilde{t}^{(1)}_{\mu}(a) &= -P^{(1)}_{\mu\mu^{\prime}}(a)r^{\mu^{\prime}}_{a}\,,&(P~\rm{wave})\\
    \tilde{t}^{(2)}_{\mu\nu}(a) &= P^{(2)}_{\mu\nu\mu^{\prime}\nu^{\prime}}(a)r^{\mu^{\prime}}_{a}r^{\nu^{\prime}}_{a}\,,&(D~\rm{wave})
\label{covariant-tensors}
\end{aligned}
\end{eqnarray}
where $r_a = p_b-p_c$. The spin factors for the $S,P$, and $D$ wave decays are
\begin{equation}
\begin{aligned}
&S_n = 1,&(S\ \rm wave)\\
&S_n = \tilde{T}^{(1)\mu}(D)\tilde{t}^{(1)}_\mu(a),&(P\ \rm wave)\\
&S_n = \tilde{T}^{(2)\mu\nu}(D)\tilde{t}^{(2)}_{\mu\nu}(a),&(D\ \rm wave)
\end{aligned}
\end{equation}
where $\tilde{T}^{(l)\mu}$ has the same definition as $\tilde{t}^{(l)\mu}$ in Ref.~\cite{covariant-tensors}. The tensor describing the $D$ meson decay is denoted by $\tilde{T}$ and that of the $a$ meson decay is denoted by $\tilde{t}$ in this paper.


\subsection{Fit results}
\begin{figure}[htbp]
  \centering
 \includegraphics[width=0.45\textwidth]{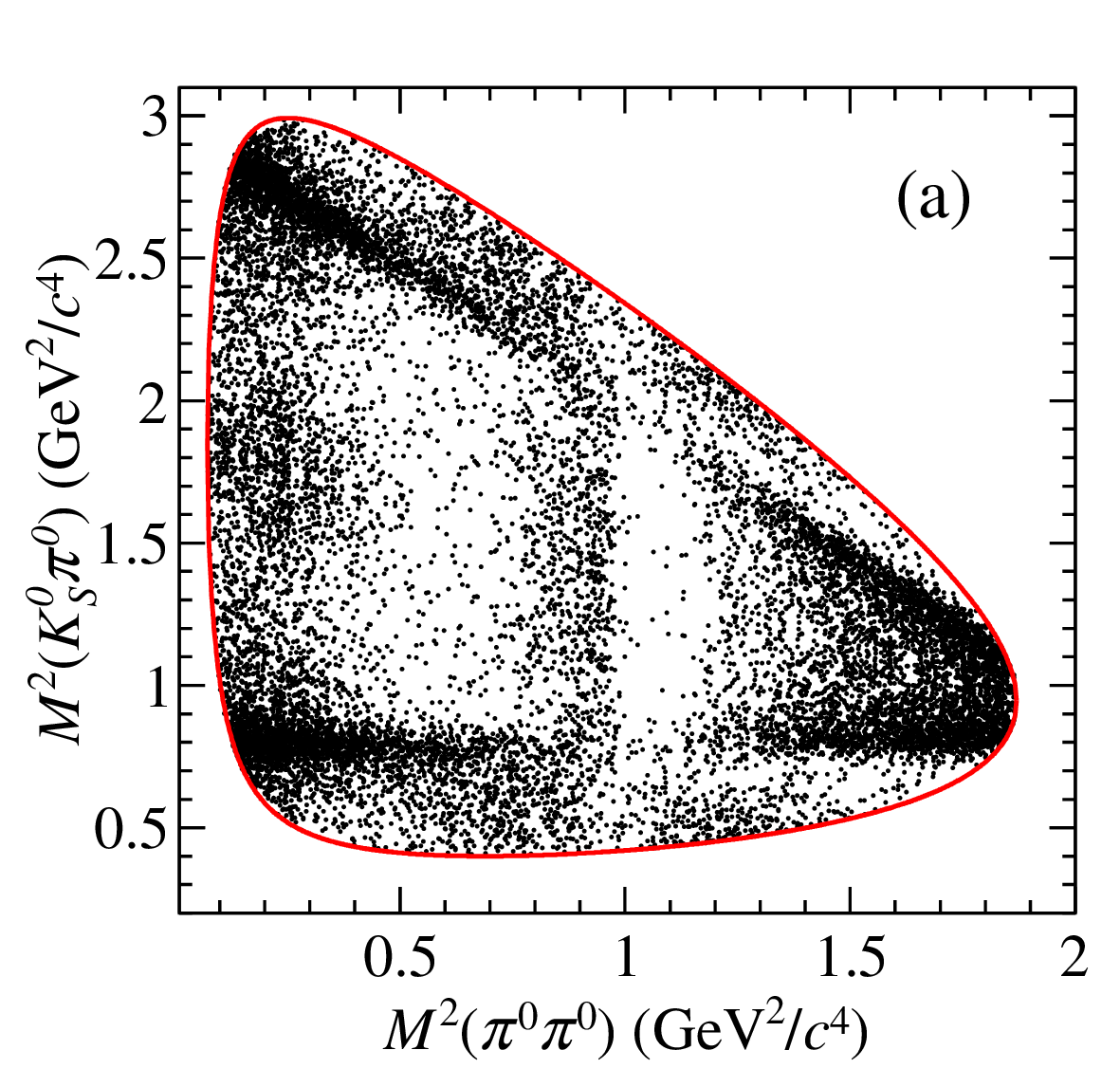}
  \includegraphics[width=0.45\textwidth]{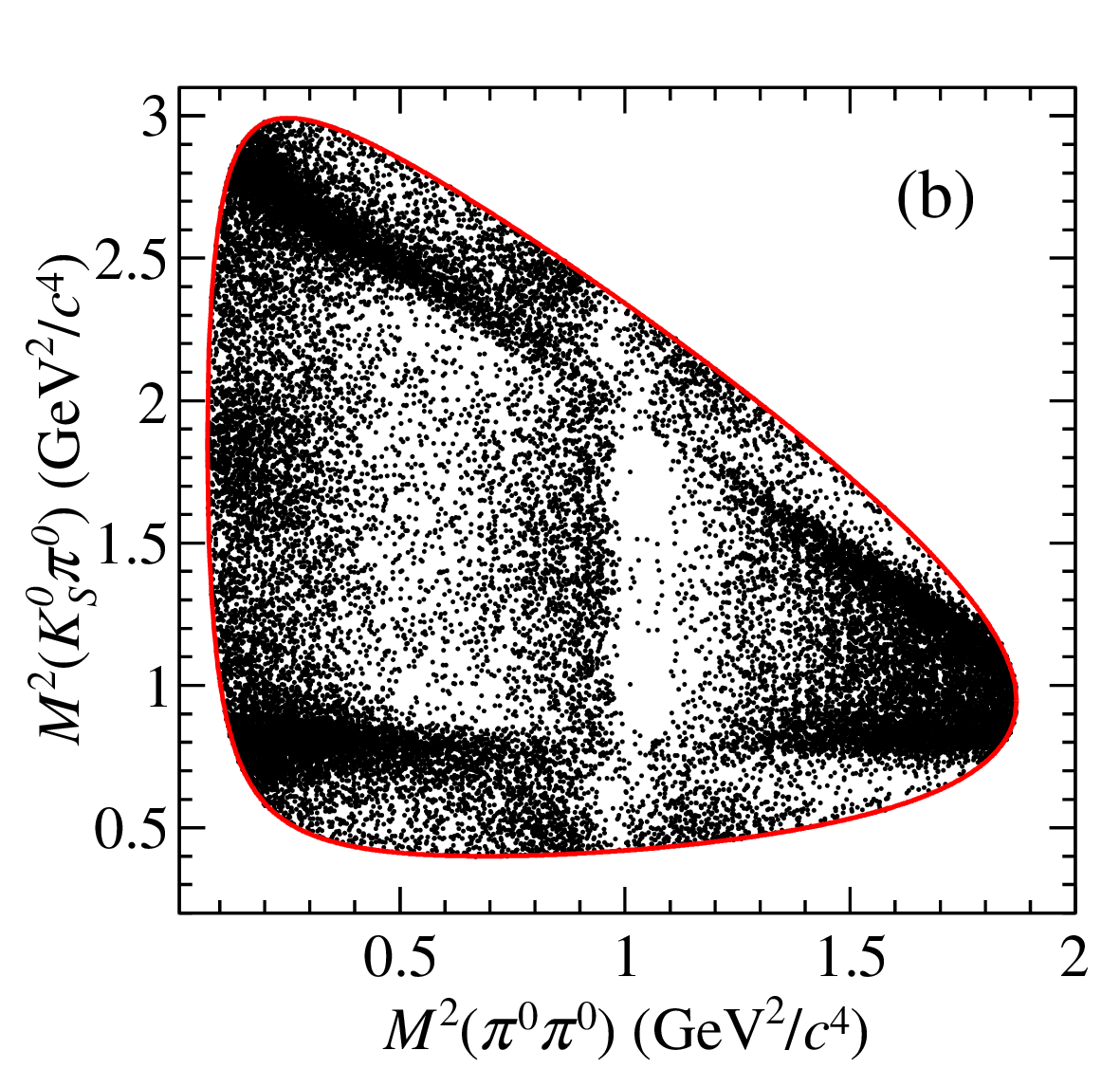}
  
  \caption{The Dalitz plots of $M^2_{K_S^0\pi^0}$  versus $M^2_{\pi^0\pi^0}$ from the selected data sample (a) and the selected signal MC sample (b) generated based on the amplitude analysis results. These plots show two entries per candidate, one for each possible $ K^0_S \pi^0 $ combination. The red curve indicates the kinematic boundary.}
  \label{fig:dalitz}
\end{figure}

The Dalitz plot of $M^2_{K^0_S \pi^0}$ versus $M^2_{\pi^0\pi^0}$ from the selected data samples is shown in the left panel of Fig.~\ref{fig:dalitz}, symmetrized for the indistinguishable $\pi^0$ candidates (two entries per candidate).

There is a clear structure caused by the $\bar K^{*}(892)^0\pi^0$ component, and its magnitude and phase are fixed to 1.0 and 0.0 in the amplitude analysis, respectively. 

Other possible processes are subsequently tested, including $\bar K^{*}(1410)^0$, $\bar K^{*}_2(1430)^0$, $\bar K^{*}(1680)^0$, $\bar K^{*}(1950)^0$, $f_2(1270)$, $(K^0_S\pi^0)_{S-{\rm wave}}$ and $(\pi^0\pi^0)_{S-{\rm wave}}$. The final choice in the nominal fit is the amplitudes of $D^0\to \bar K^{*}(892)^0\pi^0$, $D^0\to \bar K^{*}(1410)^0\pi^0$, $D^0\to \bar K_2^{*}(1430)^0\pi^0$, $D^0\to (K_S^0\pi^0)_{S\rm -wave}\pi^0$, $D^0\to \bar K^{*}(1680)^0\pi^0$, $D^0\to K_S^0(\pi^0\pi^0)_{S\rm -wave}$ and $D^0\to K_S^0f_2(1270)$, which have statistical significances greater than three standard deviations. The statistical significances are determined from the changes in log-likelihood and the numbers of degrees of freedom when the fits are performed with and without the amplitude included, compared with the nominal fit.

The fit fraction (FF) for each amplitude is defined as the ratio between the integral of its absolute square over the PHSP and the integral of the absolute square of the total amplitude. For the $n^{\rm{th}}$ amplitude, the FF is expressed as
\begin{eqnarray}\begin{aligned}
  {\rm FF}_{n} = \frac{\int\left|\rho_{n}e^{i\phi_{n}}\mathcal{A}_{n}\right|^{2} d\Phi_{3}}{\int \left|\mathcal{M}\right|^{2} d\Phi_{3}}\,.
  \label{Fit-Fraction-Definition}
\end{aligned}\end{eqnarray}
In practice, the FF is computed numerically using generator-level PHSP MC events. The discrete form of Eq.~(\ref{Fit-Fraction-Definition}) becomes 
\begin{eqnarray}\begin{aligned}
  {\rm FF}_{n} = \frac{\sum^{N_{\rm gen}} \left|\rho_{n}e^{i\phi_{n}}\mathcal{A}_{n}\right|^{2}}{\sum^{N_{\rm gen}} \left|\mathcal{M}\right|^{2}}\,, 
  \label{Fit-Fraction-Definition-MC}
\end{aligned}\end{eqnarray}
where $N_{\rm gen}$ is the number of PHSP MC events at generator level.
The sum of these FFs may not be unity if there is net constructive or destructive interference. Interference~(IN) between the $n^{\rm{th}}$ and $n^{\prime\rm{th}}$ amplitudes is defined as the ratio between the cross-term over the PHSP and the integral of the absolute square of the total amplitude:
\begin{equation}
{\rm IN}_{nn^{\prime}} = \frac{\int 2\text{Re}[\rho_{n}e^{i\phi_{n}}\mathcal{A}_{n}(\rho_{n^{\prime}}e^{i\phi_{n^{\prime}}}\mathcal{A}_{n^{\prime}})^{*}]d\Phi_3}{\int \left|\mathcal{M}\right|^{2} d\Phi_3}\,.
\label{interferenceFF-Definition}
\end{equation}
Numerical evaluation of the interference terms is performed using the discrete representation:
\begin{equation}
{\rm IN}_{nn^{\prime}} = \frac{\sum^{N_{\rm gen}} 2\text{Re}[\rho_{n}e^{i\phi_{n}}\mathcal{A}_{n}(\rho_{n^{\prime}}e^{i\phi_{n^{\prime}}}\mathcal{A}_{n^{\prime}})^{*}]}{\sum^{N_{\rm gen}} \left|\mathcal{M}\right|^{2}}\,.
\label{interferenceFF-Definition-MC}
\end{equation}
The interferences between the amplitudes are listed in Table~\ref{tab:inter} of Appendix~\ref{app:interference}.

\begin{figure}
	\centering
    \includegraphics[width=0.45\textwidth]{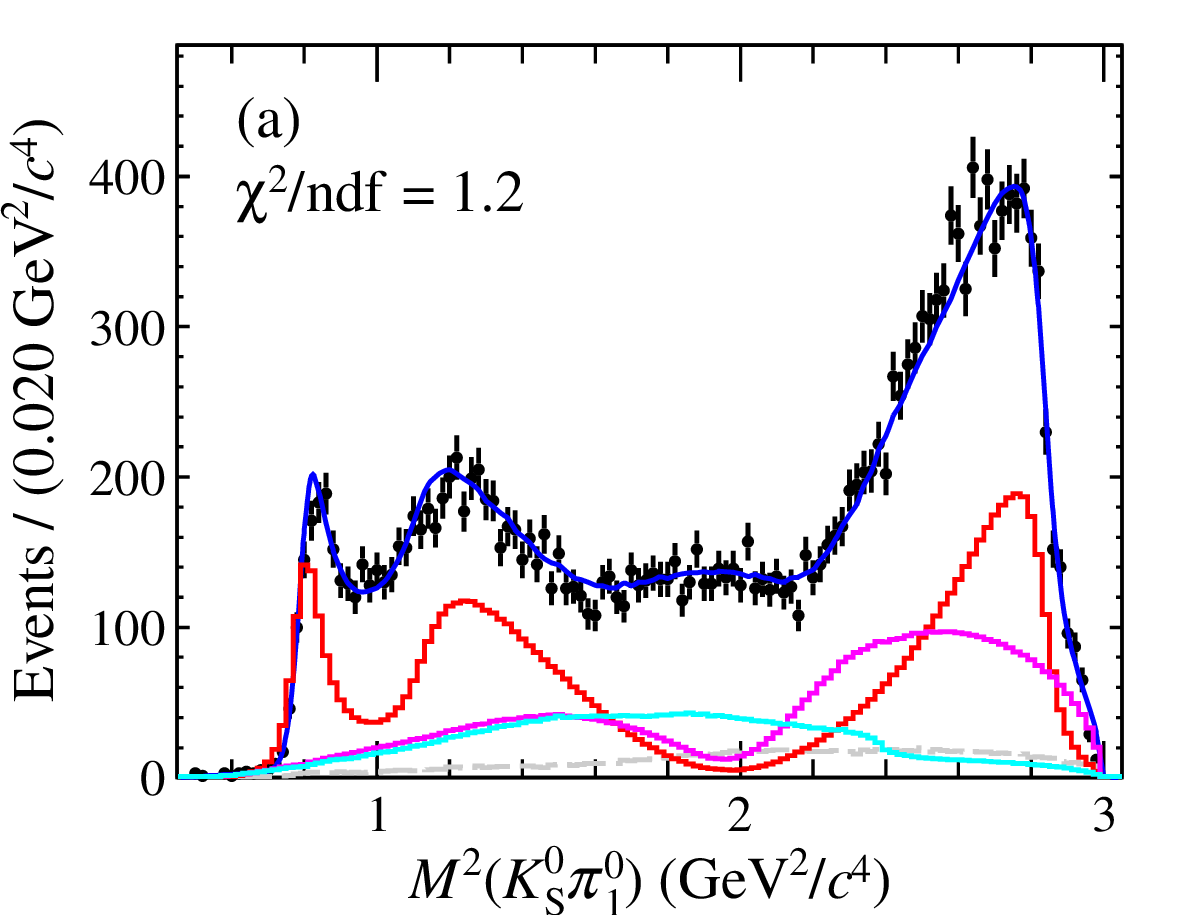}
    \includegraphics[width=0.45\textwidth]{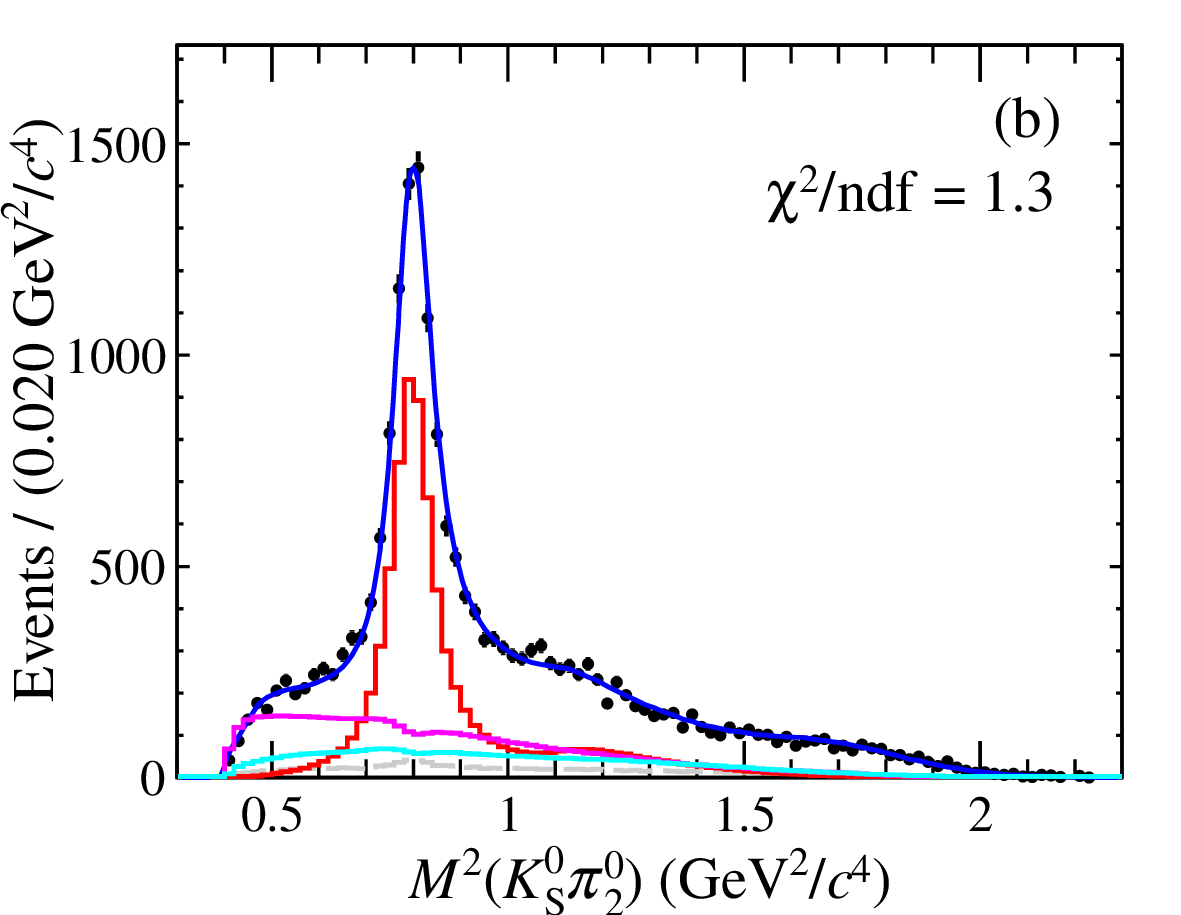}\\
    \includegraphics[width=0.45\textwidth]{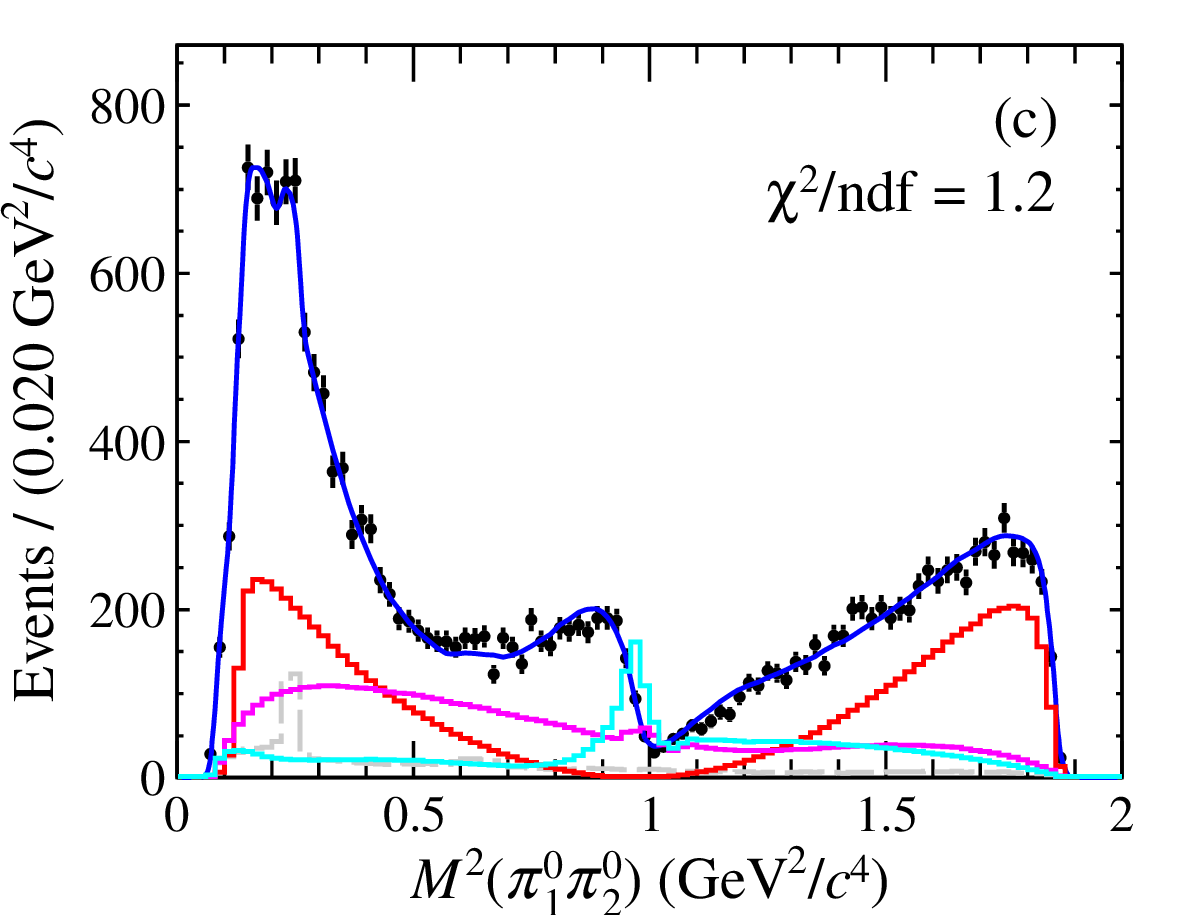}
    \includegraphics[width=0.45\textwidth]{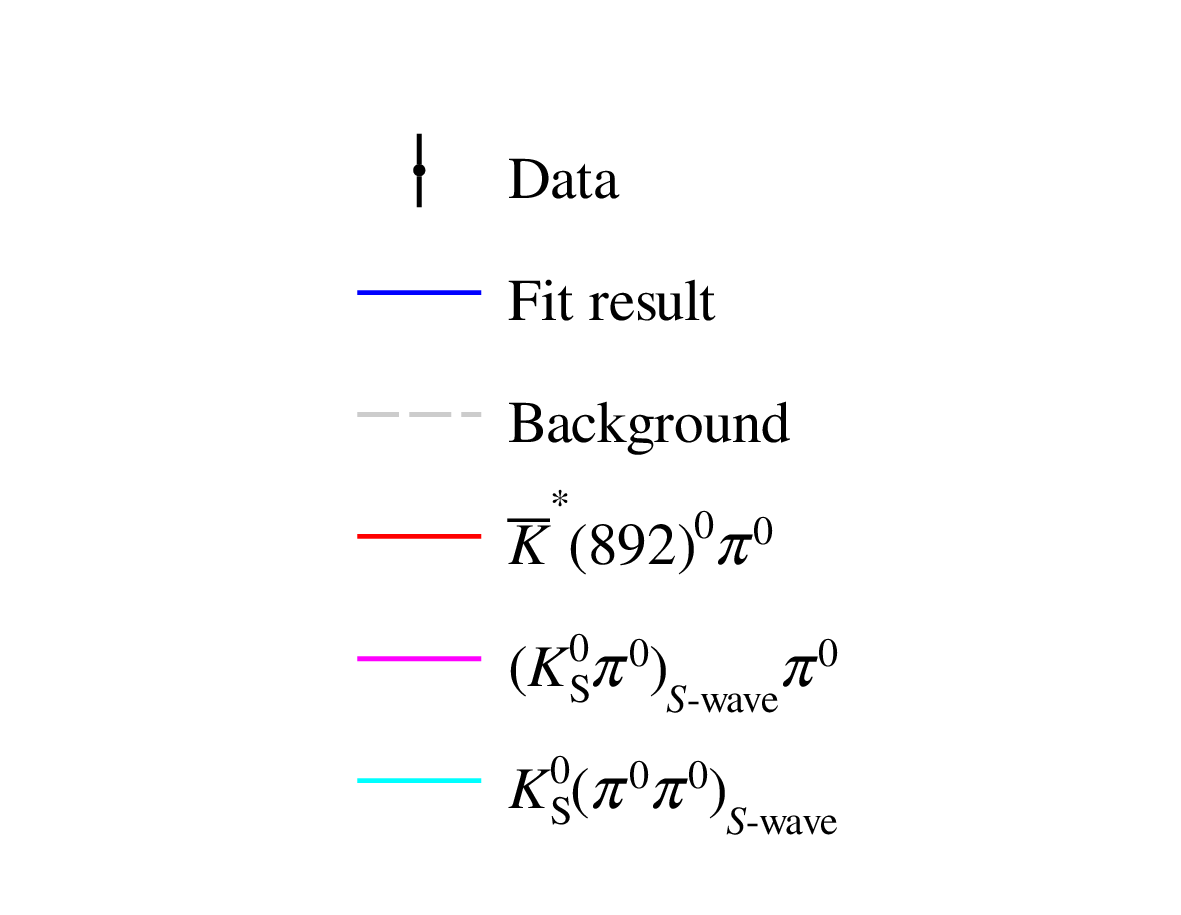}\\
	\caption{The projections of the nominal fit on (a) $M^2(K_S^0\pi^0_1)$, (b) $M^2(K_S^0\pi^0_2)$ and (c) $M^2(\pi_1^0\pi_2^0)$. The data are represented by the points with error bars, the fit results by the dark blue line, and the background by the gray dashed line. Other colored lines show the main components of the fit model. The two $\pi^0$s are distinguished by momentum,  where the momentum of $\pi^0_1$ is higher than $\pi^0_2$.}

    \label{fig:fitresult}
\end{figure}

It is impractical to analytically propagate the uncertainties of the magnitudes and phases to the FF. Instead, the variables are randomly varied 500 times based on their covariance matrix obtained from the fit, and in each iteration, the FFs are calculated to determine the statistical uncertainties. 
A Gaussian function is subsequently used to fit the distribution of each FF. The width of this function is assigned as the uncertainty of the corresponding FF. The magnitudes, phases and FFs for different amplitudes are listed in Table~\ref{tab:fitresult}. The Dalitz plot of the signal MC sample generated based on the result of the amplitude analysis is shown in the right panel of Fig.~\ref{fig:dalitz}. The mass projections of the nominal fit are shown in Fig.~\ref{fig:fitresult}.

\begin{table}[htbp]
  \centering
  \begin{tabular}{|l r@{$\,\pm\,$}l r@{$\,\pm\,$}l@{$\,\pm\,$}l r@{$\,\pm\,$}l@{$\,\pm\,$}l|}
    \hline
    Amplitude &\multicolumn{2}{c}{Magnitude} &\multicolumn{3}{c}{Phase~(rad)} &\multicolumn{3}{c|}{FF(\%)} \\
    \hline	
    $D^0\to \bar K^{*}(892)^0\pi^0$ &\multicolumn{2}{c}{1~(fixed)}
    &\multicolumn{3}{c}{0~(fixed)}    &40.0  &0.9 &0.6 \\
    $D^0\to \bar K^{*}(1410)^0\pi^0$ &\multicolumn{2}{c}{0.33$\,\pm\,$0.09}  &3.74 &0.22 &0.31    &0.5  &0.3 &0.4 \\
    $D^0\to \bar K_2^{*}(1430)^0\pi^0$
    &\multicolumn{2}{c}{0.84$\,\pm\,$0.07}  &5.57 &0.11 &0.06      &1.6 &0.3 &0.2 \\
    $D^0\to \bar K^{*}(1680)^0\pi^0$
    &\multicolumn{2}{c}{1.56$\,\pm\,$0.29}  &1.73 &0.18 &0.21   &3.7  &1.5 &2.5   \\
    $D^0\to K_S^0f_2(1270)$
    &\multicolumn{2}{c}{2.04$\,\pm\,$0.22}  &1.02 &0.13 &0.05  & 2.5  &0.6 &0.5   \\
    $D^0\to (K_S^0\pi^0)_{S\rm-wave}\pi^0$
    &\multicolumn{2}{c}{--}
    &\multicolumn{3}{c}{--}    &30.4 &5.4 &2.2  \\
    ${\alpha}_{K\pi}$
    &\multicolumn{2}{c}{2.86$\,\pm\,$0.24$\,\pm\,$0.12}  &8.57 &0.16 &0.31  &\multicolumn{3}{c|}{ }  \\
    ${\alpha}_{K\eta^{\prime}}$
    &\multicolumn{2}{c}{21.15$\,\pm\,$2.27$\,\pm\,$2.43}  &$-10.77$ &0.14 &0.23  &\multicolumn{3}{c|}{ }  \\
    ${\alpha}_{32}$
    &\multicolumn{2}{c}{4.10$\,\pm\,$0.53$\,\pm\,$0.90}  &3.93 &0.09 &0.06  &\multicolumn{3}{c|}{ }  \\
    $D^0\to K_S^0(\pi^0\pi^0)_{S-{\rm wave}}$
    &\multicolumn{2}{c}{0.77$\,\pm\,$0.04}  &0.49 &0.06 &0.06     &14.6  &1.7 &0.6 \\
    $\beta_1$
    &\multicolumn{2}{c}{8.71$\,\pm\,$0.42$\,\pm\,$0.59}  &1.17 &0.04 &0.02  &\multicolumn{3}{c|}{}  \\
    $f_{11}^{\rm prod}$
    &\multicolumn{2}{c}{8.07$\,\pm\,$0.23$\,\pm\,$0.25}  &$-2.13$ &0.03 &0.03  &\multicolumn{3}{c|}{}  \\
    \hline
    Total &\multicolumn{2}{c}{ }
    &\multicolumn{3}{c}{ }  
    &\multicolumn{3}{c|}{93.3}  \\
    \hline
  \end{tabular}
  \caption{Magnitudes, phases and FFs for different amplitudes in $D^0\to K^0_S\pi^0\pi^0$. The uncertainties in magnitudes are statistical only. The first and second uncertainties for the phases and FFs are statistical and systematic, respectively.} 
  \label{tab:fitresult}
\end{table}

\subsection{Systematic uncertainties in the amplitude analysis}
\label{sec:PWA-Sys}
The systematic uncertainties in the amplitude analysis are described below and summarized in Table~\ref{tab:sys}. 
\begin{itemize}
\item[\uppercase\expandafter{\romannumeral1}] 
Amplitude model: \\
The masses and widths of resonances are varied to their $\pm 1\sigma$ boundaries as tabulated in~\cite{PDG}, with two dedicated refits performed to evaluate the corresponding systematic uncertainties.
The $\pi^0 \pi^0$ $S$-wave $K$-matrix formalism is modified according to the parameters $f_{1j}^{\mathrm{prod}}$, $\beta_{\alpha}$ at their $\pm 1\sigma$ ranges as given in~\cite{KPsnew2}. For $\pi^0 \pi^0$ $S$-wave $K$-matrix formalism, two independent fits are performed by setting the parameters to their upper and lower $1\sigma$ limits. The systematic uncertainty for every parameter is determined by its maximum deviation between these boundary fits and the central result. The total amplitude model uncertainty combines all individual contributions in quadrature.

\item[\uppercase\expandafter{\romannumeral2}]
Effective radius: \\
The systematic uncertainty on the $R$ parameters in the Blatt–Weisskopf factors is obtained by repeating the fit, with the effective radii of the intermediate states and the $D^0$ meson varied independently by $\pm 0.5 ({\text{GeV}/c})^{-1}$ about their corresponding central values.

\item[\uppercase\expandafter{\romannumeral3}]  
Background: \\
In analyzing the impact of background on the amplitude model, there are two primary sources of influence: background size and background shape.  
The background size is associated with the signal purity ($w_{\rm{sig}}$) in Eq.~(\ref{likelihood3}). 
The systematic uncertainty from the $w_{\rm{sig}}$ parameter is evaluated by performing two additional fits with the parameter fixed at its $\pm 1 \sigma $ boundaries. The corresponding systematic uncertainty for each observable is then taken as the maximum deviation between these boundary fits and the nominal result. The background shape is related to the background function ($B(p_j)$) in Eq.~(\ref{pdf:bkg}). An alternative background sample is used to determine the background shape, where the relative fractions of background processes from direct $\bar{q} q$ are varied by the statistical uncertainties of the known cross sections~\cite{Luminosity}. The square root of the quadratic sum of these two uncertainties is taken as the background uncertainty.

\item[\uppercase\expandafter{\romannumeral4}] 
Experimental effects: \\
To estimate the uncertainties associated with $\gamma_{\epsilon}$, as defined in Eq.~(\ref{pwa:gamma}), the amplitude model is refitted  by varying reconstruction efficiencies of $K_S^0$ and $\pi^0$ according to their uncertainties. The maximum deviations between the nominal and refitted parameters are then taken as the systematic uncertainties.
\item[\uppercase\expandafter{\romannumeral5}] 
Fit bias:\\
To study the possible bias from the fit procedure, an ensemble of 600 signal MC samples are generated according to the results of the amplitude analysis. The fit procedure is repeated for each signal MC sample, and the pull distributions of the amplitude results are fitted by a Gaussian. Finally, the FFs and phases of all resonances, as well as their statistical uncertainties, are corrected by the fitted mean values of the pull distribution, and the uncertainty of the fitted mean values is assigned as the corresponding systematic uncertainties. 

\item[\uppercase\expandafter{\romannumeral6}] 
Insignificant amplitudes:\\
With all possible resonances accounted for, the related systematic uncertainty is neglected.

\end{itemize}

\begin{table}[htbp]
    \centering
    \begin{tabular}{|lccccccc|}
    \hline
    \   &\multicolumn{7}{c|}{Source}   \\
    \hline
    Amplitude  &\  &I  &II  &III   &IV   &V   &Total\\
    \hline
    $D^0\to \bar K^{*}(892)^0\pi^0$
    &FF     &0.60 &0.34 &0.22 &0.00 &0.04 &0.72\\
    \multirow{2}{*}{$D^0\to \bar K^{*}(1410)^0\pi^0$}
    &FF     &1.26 &0.31 &0.10 &0.10 &0.04 &1.31\\ 
    &$\phi$ &1.38 &0.07 &0.18 &0.05 &0.04 &1.39 \\
    \multirow{2}{*}{$D^0\to \bar K^{*}_2(1430)^0\pi^0$}
    &FF     &0.43 &0.49 &0.20 &0.03 &0.04 &0.68\\ 
    &$\phi$ &0.41 &0.33 &0.10 &0.04 &0.04 &0.54 \\
    $D^0\to ({K^0_S\pi^0})_{S\rm -wave}\pi^0$
    &FF     &0.30 &0.23 &0.15 &0.01 &0.04 &0.41 \\ 
    \multirow{2}{*}{$D^0\to \bar K^{*}(1680)^0\pi^0$}
    &FF     &1.66 &0.20 &0.11 &0.10 &0.04 &1.68 \\ 
    &$\phi$ &1.06 &0.40 &0.06 &0.03 &0.04 &1.14 \\
    \multirow{2}{*}{$D^0 \to K^0_S(\pi^0\pi^0)_{S\rm -wave}$}
    &FF     &0.18 &0.25 &0.10 &0.02 &0.04 &0.33 \\ 
    &$\phi$ &0.94 &0.04 &0.10 &0.16 &0.04 &0.96 \\
    \multirow{2}{*}{$D^0 \to K^0_Sf_2(1270)$}
    &FF     &0.62 &0.44 &0.13 &0.07 &0.05 &0.78 \\ 
    &$\phi$ &0.37 &0.06 &0.13 &0.06 &0.04 &0.40 \\
    \hline
    \multirow{2}{*}{${\alpha}_{K\pi}$} 
    &$\rho$ &0.37 &0.33 &0.08 &0.13 &0.05 &0.52 \\ 
    &$\phi$ &1.03 &0.05 &0.10 &0.27 &0.05 &1.07 \\
    \multirow{2}{*}{${\alpha}_{K\eta^{\prime}}$} 
    &$\rho$ &1.82 &0.04 &0.14 &0.01 &0.04 &1.83 \\ 
    &$\phi$ &1.95 &0.13 &0.19 &0.11 &0.06 &1.97 \\
    \multirow{2}{*}{${\alpha}_{32}$} 
    &$\rho$ &1.67 &0.13 &0.09 &0.05 &0.06 &1.68 \\ 
    &$\phi$ &0.43 &0.18 &0.10 &0.37 &0.04 &0.60 \\
    \hline
    \multirow{2}{*}{$\beta_{1}$} 
    &$\rho$ &1.38 &0.19 &0.11 &0.13 &0.04 &1.40 \\ 
    &$\phi$ &0.41 &0.31 &0.10 &0.21 &0.04 &0.57 \\
    \multirow{2}{*}{$f_{11}^{\rm prod}$} 
    &$\rho$ &1.06 &0.27 &0.13 &0.15 &0.04 &1.11 \\ 
    &$\phi$ &0.94 &0.16 &0.09 &0.06 &0.04 &0.96 \\
    \hline
    \end{tabular}
    \caption{Systematic uncertainties on the magnitudes $(\rho)$, phases $(\phi)$ and FFs for the different components in the amplitude model, expressed as ratios to their statistical uncertainties. (I) Amplitude model, (II) Effective radius, (III) Background, (IV) Experimental effects, and (V) Fit bias.}
    \label{tab:sys}
\end{table}
\section{Branching fraction measurement}
The BF of the  $D^0 \to K^0_S\pi^0\pi^0$ decay is measured with the DT technique applying the same tag modes as those utilized in the amplitude analysis.
The selection criteria follow those discussed in Sec.~\ref{ST-selection}, which are identical to the criteria used in the amplitude analysis, except for the specific requirements listed in Sec.~\ref{sec:pwa-select}.

For each ST mode, the following relations are established~\cite{QCBF}:
\begin{equation}
  N_{\text{tag}}^{\text{ST}} = 2N_{D^{0} \bar D^{0}}\cdot \mathcal{B}_{\text{tag}}\cdot\epsilon_{\text{tag}}^{\text{ST}}\cdot(1+y_D^2)\cdot(1+r_{\rm tag}^2-2r_{\rm tag} R_{\rm tag} y_D \cos\!\delta_{\rm tag})\,, \label{eq-ST}
\end{equation}
\begin{equation}
\begin{split}
N_{\text{tag,sig}}^{\text{DT}} = & 2N_{D^{0} \bar D^{0}}\cdot \mathcal{B}_{\rm sub}\cdot \mathcal{B}_{\text{tag}}\cdot\mathcal{B}_{\text{sig}}\cdot \epsilon_{\text{tag,sig}}^{\text{DT}} \\
& \cdot(1+y_D^2)\cdot[1+r_{\rm tag}^2-2r_{\rm tag} R_{\rm tag} \cos\!\delta_{\rm tag}(2F_{+}^{\rm sig}-1)]\,,
\end{split}
\label{eq-DT}
\end{equation}
where $N_{\text{tag}}^{\text{ST}}$ is the ST yield for a specific tag mode, $N_{D^{0} \bar D^{0}}$ is the total number of $D^{0} \bar D^{0}$ pairs produced from $e^{+}e^{-}$ collisions, $\mathcal{B}_{\text{tag}}$ is the BF of the tag mode, and $\epsilon_{\text{tag}}^{\text{ST}}$ is the ST efficiency for the tag mode. The observable $N_{\text{tag,sig}}^{\text{DT}}$ is the DT yield, $\mathcal{B}_{\text{sig}}$ is the BF of the signal mode, and $\epsilon_{\text{tag,sig}}^{\text{DT}}$ is the efficiency for simultaneously reconstructing the signal and specific tag mode. To account for the reconstruction of the signal through subsequent decays, the factor $\mathcal{B}_{\text{sub}} = \mathcal{B}(K^0_S \to \pi^+\pi^-) \mathcal{B}^2(\pi^0 \to \gamma\gamma)$ is introduced. Additionally, $y_D$ is the $D^0-\bar D^0$ mixing parameter, and $F_+$ is the $\mathit{CP}$-even fraction of the signal decay.
For the $D^0 \to K_S^0\pi^0\pi^0$ decay, which is a pure $\mathit{CP}$-even decay, $F_+$ is equal to unity. The parameters $r$, $R$, and $\delta$ are introduced to account for quantum-correlation effects, and their values for the three tag modes are listed in Table~\ref{tab:qcc}. Combining the two equations above and ignoring the term $2r_{\rm tag} R_{\rm tag} y_D \cos\!\delta_{\rm tag}$, the absolute BF of  $D^0 \to K^0_S\pi^0\pi^0$ is determined by
\begin{equation}
  \mathcal{B}_{\text{sig}} = \frac{N_{\text{tag,sig}}^{\text{DT}}}{\begin{matrix} \mathcal{B}_{\rm sub}\cdot N_{\text{tag}}^{\text{ST}}\cdot \epsilon^{\text{DT}}_{\text{tag,sig}}/\epsilon_{\text{tag}}^{\text{ST}}\cdot[1-\frac{2r_{\rm tag} R_{\rm tag} \cos\!\delta_{\rm tag}}{1+r_{\rm tag}^2}(2F_{+}^{\rm sig}-1)]\end{matrix}}\,.\label{BR-formula}
\end{equation} 

\begin{table}[hbtp]
  \begin{center}
    
    \begin{tabular}{|l r@{ $\pm$ }l r@{ $\pm$ }l r@{ $\pm$ }l |}
      \hline
      Tag mode &\multicolumn{2}{c}{$r~(\%)$}  &\multicolumn{2}{c}{$R$}  &\multicolumn{2}{c|}{$\delta~(^{\circ})$} \\ 
      \hline
      $\bar {D}^0\to K^+\pi^-$	    &\multicolumn{2}{c}{$5.855^{+0.009}_{-0.010}$\cite{QCC1}} 
      &\multicolumn{2}{c}{1} 
      &\multicolumn{2}{c|}{$191.4 \pm 2.4$\cite{QCC1}}   \\
      $\bar {D}^0\to K^+\pi^-\pi^0$   &4.41 &0.11\cite{QCC2} 
      &0.79 &0.04\cite{QCC2}
      &196 &11\cite{QCC2} \\
      $\bar {D}^0\to K^+\pi^-\pi^-\pi^+$ &5.50 &0.07\cite{QCC2} 
      &\multicolumn{2}{c}{$0.44^{+0.09}_{-0.10}$\cite{QCC2}}
      &\multicolumn{2}{c|}{$161^{+28}_{-18}$\cite{QCC2}}\\
      \hline
    \end{tabular}
    \caption{ The input values of $r$, $R$, $\delta$ for the three tag modes.}
    \label{tab:qcc}
  \end{center}
\end{table}

\begin{figure}[htbp]
  \centering
  \includegraphics[width=0.45\textwidth]{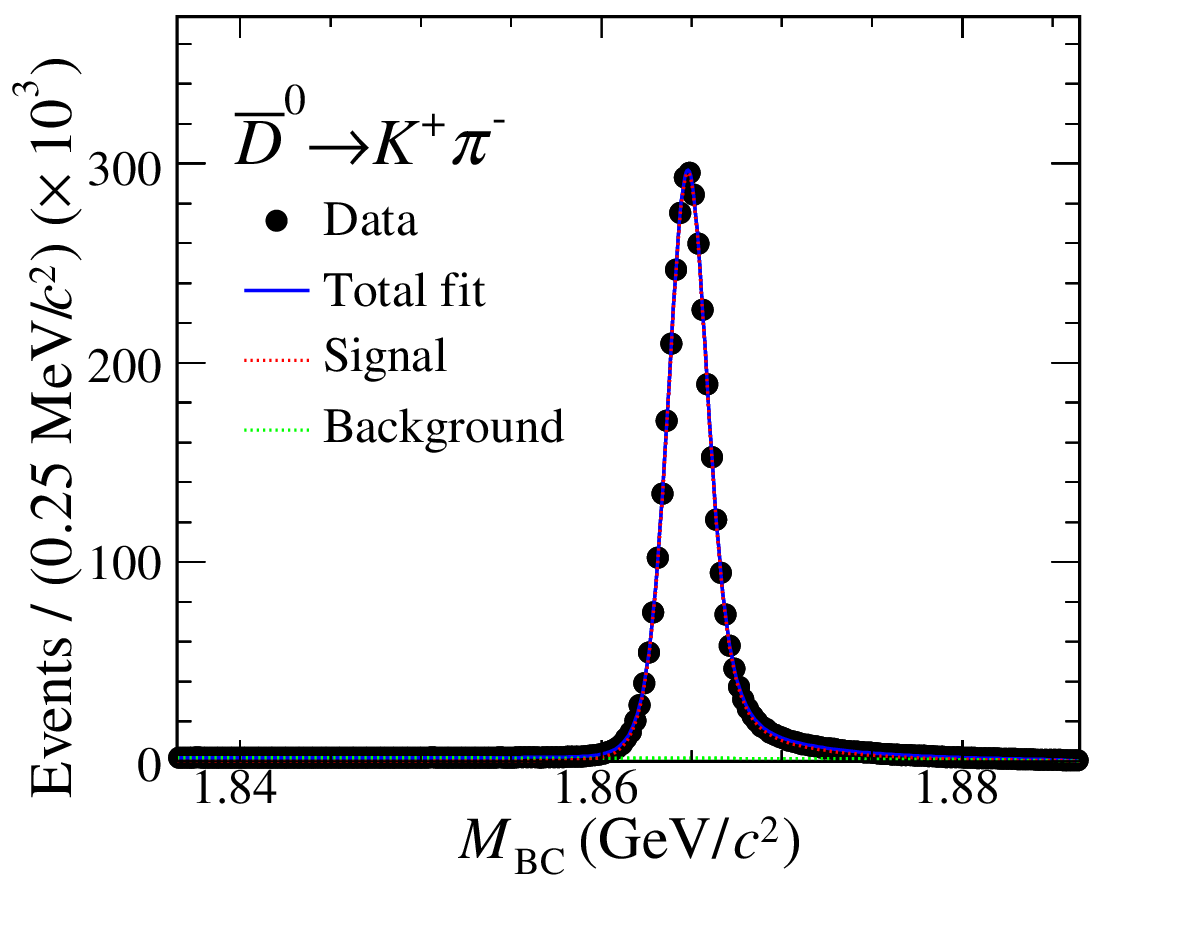}
  \includegraphics[width=0.45\textwidth]{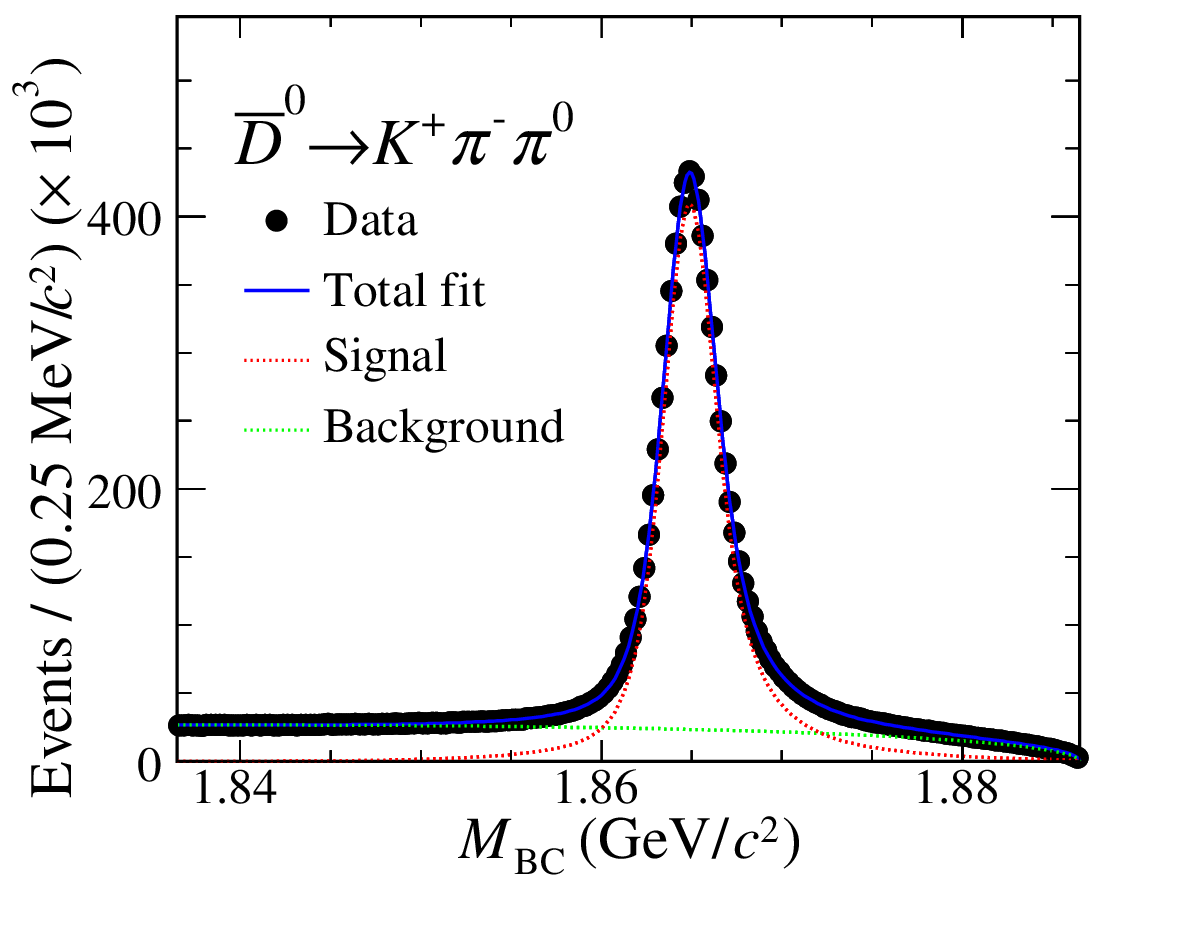}
  \includegraphics[width=0.45\textwidth]{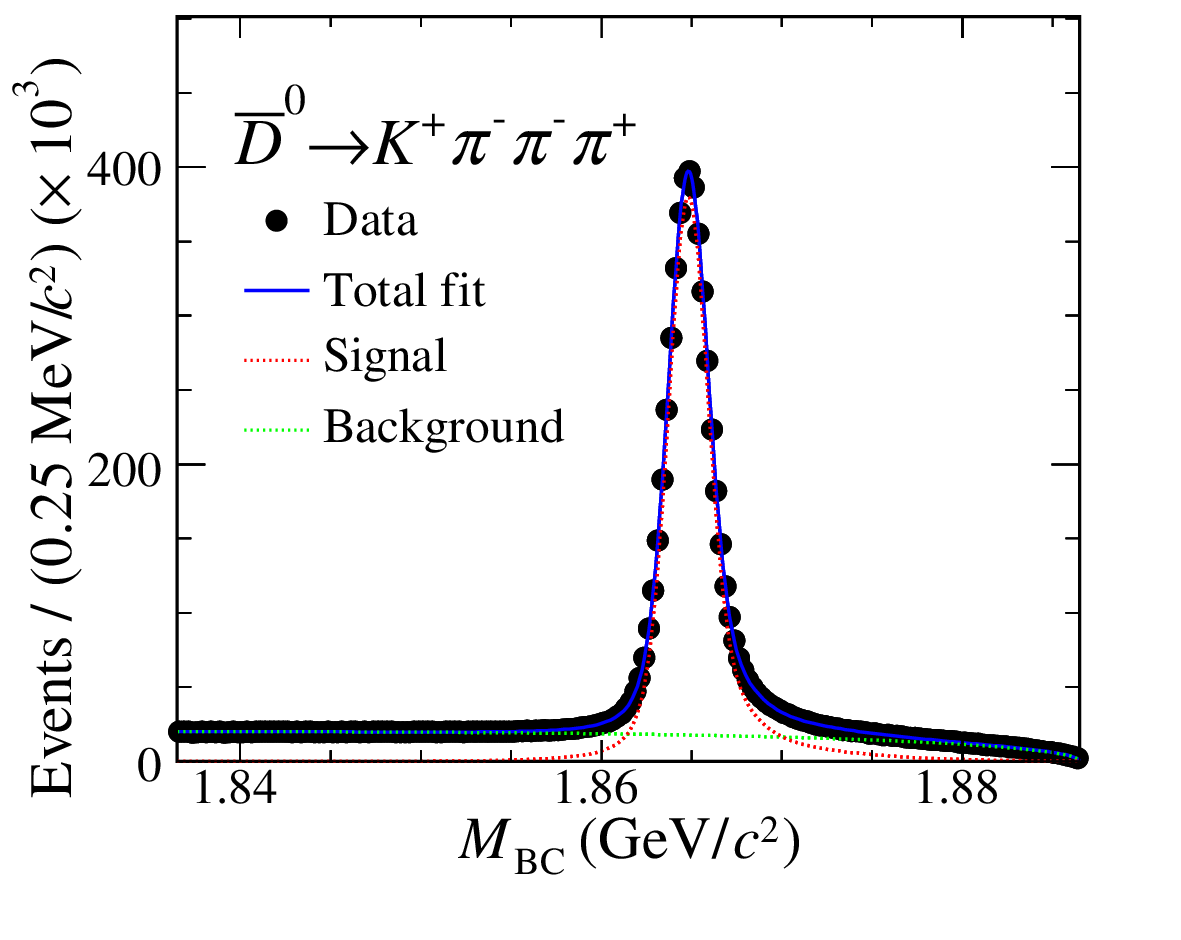}
  \caption{Fits to the $M_{\rm BC}$ distributions of the ST candidates for $\bar D^0 \to K^+ \pi^-$~(a), $\bar D^0 \to K^+ \pi^- \pi^0$~(b) and $\bar D^0 \to K^+ \pi^- \pi^- \pi^+ $~(c). The points with error bars are data. The blue curves are the fit projections. The dotted red curves are the signal. The dotted green curves are the fitted combinatorial background shapes.}
  \label{fig:ST_yield}
\end{figure}

The value of $N_{\text{tag}}^{\text{ST}}$ is obtained from a one-dimensional (1D) binned fit to the $M_{\rm BC}$ distribution and the peaking backgrounds have been subtracted, as shown in Fig.~\ref{fig:ST_yield}. 
The signal shape is modeled by an MC-simulated shape convolved with a double-Gaussian function describing the resolution difference between data and MC simulation, and the background shape is described by an ARGUS function~\cite{ARGUS}. The corresponding $\epsilon_{\text{tag}}^{\text{ST}}$ is estimated with the inclusive MC sample, where the peaking backgrounds have been removed from the samples.

The total DT yield from all three tag modes is determined to be $N_{\text{tag,sig}}^{\text{DT}} = 20865 \pm 166$ via a 2D fit to the distribution of $M_{\rm BC}^{\rm tag}$ versus $M_{\rm BC}^{\rm sig}$.
The fit includes dedicated PDFs to model the peaking backgrounds arising from the ST side. The PDFs of the 2D fit are the same as those in Sec~\ref{sec:pwa-select}. 
$\epsilon^{\text{DT}}_{\text{tag,sig}}$ is determined with the signal MC sample in which the $D^0 \to K^0_S\pi^0\pi^0$ events are generated according to the result of the amplitude analysis.
The values of these efficiencies are summarized in Table~\ref{ST-eff}.

\begin{table}[htbp]
    \begin{center}
      \begin{tabular}{|lccc|}
        \hline
        Tag mode         &$N^{\text{ST}}_{\text{tag}} (\times 10^{3})$  & $\epsilon^{\text{ST}}_{\text{tag}}(\%)$  &$\epsilon^{\text{DT}}_{\text{tag,sig}}(\%)$ \\
        \hline
	$\bar {D}^0\to K^+\pi^-$              &$3820.8 \pm 2.0$  &$ 66.67 \pm 0.01$  &$13.112 \pm 0.007$   \\  
	$\bar {D}^0\to K^+\pi^-\pi^0$         &$7926.5 \pm 3.3$  &$ 37.92 \pm 0.01$  &$~6.512 \pm 0.003$        \\  
	$\bar {D}^0\to K^+\pi^-\pi^-\pi^+$    &$5140.1 \pm 2.6$  &$ 42.20 \pm 0.01$  &$~7.256 \pm 0.004$         \\  
        \hline
      \end{tabular}
    \end{center}
      \caption{The ST yields ($N^{\text{ST}}_{\text{tag}}$), ST efficiency ($\epsilon^{\text{ST}}_{\text{tag}}$) and DT efficiency ($\epsilon^{\text{DT}}_{\text{tag,sig}}$). The efficiencies do not include the branching fractions (BFs) for $\pi^0 \to \gamma\gamma$ and $K^0_S \to \pi^+\pi^-$. The uncertainties are statistical only.}
\label{ST-eff}
\end{table}

The systematic uncertainties for the BF measurement are described below and  summarized in Table~\ref{BF-Sys}.


\begin{itemize}
\item ST $\bar D^{0}$ candidates:

The uncertainty in the yield of ST $\bar D^0$ mesons is assigned to be 0.3\% by varying the signal shape, background shape, and varying the parameters of the Gaussian in the fit.


\item Tracking:

The tracking efficiency of $\pi^{\pm}$ is investigated with the DT hadronic $D\bar{D}$ events of the decays $D^0\to K^-\pi^+$, $K^-\pi^+\pi^0$, $K^-\pi^+\pi^+\pi^-$ versus $\bar{D}^0\to K^+\pi^-$, $K^+\pi^-\pi^0$, $K^+\pi^-\pi^-\pi^+$, and $D^+\to K^-\pi^+\pi^+$ versus $D^- \to K^+\pi^-\pi^-$.
The data-MC efficiency ratio for pion tracking is found to be 0.996 $\pm$ 0.001. After applying this correction factor to the MC efficiency for each pion, the statistical uncertainty of the correction factor is propagated as the systematic uncertainty of each pion. This results in a systematic uncertainty of 0.2\% on the overall pion tracking efficiency. 

\item $K_S^{0}$ reconstruction:

The data-MC efficiency ratio for $K_S^0$ reconstruction is $0.995 \pm 0.002$, which is measured with the samples of $D^0$ or $D^+$ decaying into $K_S^0\pi^+\pi^-, K_S^0 \pi^+ \pi^- \pi^0$, $K_S^0 \pi^0$, $K_S^0\pi^-, K_S^0\pi^-\pi^0$ or $K_S^0\pi^+\pi^-\pi^-$ hadronic decays. After correcting the efficiency of $K_S^0$ reconstruction by this factor, the associated systematic uncertainty is assigned as 0.2\%.

\item $\pi^{0}$ reconstruction:

The data-MC efficiency ratio for each $\pi^0$ reconstruction is $0.974 \pm 0.002$, which is measured with the samples of $D^0 \to K^-\pi^+ \pi^0$ versus $\bar{D}^0 \to K^+\pi^-$ and $\bar{D}^0 \to K^+\pi^-\pi^- \pi^+$ hadronic decays. After applying this factor to correct the efficiency of each $\pi^0$ reconstruction, the statistical uncertainty of the correction factor is propagated to the systematic uncertainty. Since the analysis involves two $\pi^0$  mesons, this results in a total systematic uncertainty of 0.4\% on the overall reconstruction efficiency.

\item MC sample size:

The uncertainty of limited MC sample size is obtained by $\sqrt{ \begin{matrix} \sum_{i} (f_{i}\frac{\delta_{\epsilon_{i}}}{\epsilon_{i}}\end{matrix})^2}$, where $f_{i}$ is the tag yield fraction, $\epsilon_{i}$ is the signal efficiency and $\delta_{\epsilon_i}$ is the uncertainty of signal efficiency of tag mode $i$. The corresponding uncertainty is determined to be 0.1\%.

\item Amplitude  model:

The uncertainty from the amplitude model is determined by varying the amplitude model parameters based on their error matrix 600 times. A Gaussian function is used to fit the distribution of 600 DT efficiencies and the fitted width  divided by the mean value is taken as an uncertainty, which is 0.1\%. 

\item BFs of subsequent decays:

In this measurement, the BFs of the daughter particles are quoted from the PDG~\cite{PDG}, which are $\mathcal{B}(K_S^0 \to \pi^+ \pi^-)=(69.20\pm 0.05)\%$ and $\mathcal{B}(\pi^0 \to \gamma \gamma)=(98.823\pm 0.034)\%$. 
Based on these values, the $\mathcal{B}_{\text{sub}}$ in Eq.~(\ref{BR-formula}) is calculated to be  $(67.58\pm0.06)\%$.
The associated uncertainty is assigned to be 0.1\% of the total BF.   

\item 2D fit:

The signal and background shapes, as well as the estimation of the size of the peaking background, are potential sources of uncertainty from the 2D fit. The mean and width of the convolved Gaussian function are varied by $\pm 1\sigma$ for the signal shape and the $q\bar q$ component in the inclusive MC sample are varied by the statistical uncertainty of the known cross section \cite{Luminosity} for the background shape. For the peaking backgrounds $D^0 \to K^0_S K^0_S$ and $D^0 \to \pi^+ \pi^- \pi^0 \pi^0$, the quoted BFs of their decays are varied by $\pm 1\sigma$. The quadratic sum of the relative BF changes, 0.5\%, is assigned as the systematic uncertainty for the 2D fit.

\item $\Delta E_{\rm sig}$ requirement:

The possible difference between data and MC simulation is accounted for by examining the $\Delta E_{\text{sig}}$ cut efficiency after applying a single-Gaussian smearing to the signal MC sample. The observed efficiency variation, $0.1\%$ is taken as the systematic uncertainty.

\item Quantum correlation correction:

The uncertainties associated with the quantum-correlation parameters $r$, $R$ and $\delta$ are propagated according to the results of Refs.~\cite{QCC1,QCC2}, resulting in a relative uncertainty of 0.5\%.

\end{itemize}

After correcting for the differences in $\pi^{\pm}$ tracking, $K_S^0$ reconstruction and $\pi^0$ reconstruction efficiencies between data and MC simulation, the BF of $D^0 \to K^0_S\pi^0\pi^0$ is determined to be $\mathcal{B}(D^0 \to K^0_S\pi^0\pi^0)$ = \BF. 

\begin{table}[htbp]
  \begin{center}
    \begin{tabular}{|lc|}
      \hline
      Source   &Uncertainty (\%)\\
      \hline
      ST $\bar D^{0}$ candidates       & 0.3 \\
      Tracking                    & 0.2 \\
      $K_S^0$ reconstruction      & 0.2\\
      $\pi^0$ reconstruction      & 0.4 \\
      MC sample size              & 0.1\\
      Amplitude  model            & 0.2 \\
      BFs of subsequent decays    & 0.1\\
      2D fit   		            & 0.5\\
      $\Delta E_{\rm sig}$ requirement	          & 0.1\\
      Quantum correlation correction	          & 0.5\\
      \hline
      Total                       & 0.9 \\
      \hline
    \end{tabular}
  \end{center}
  \caption{Relative systematic uncertainties in the BF measurement.}
   \label{BF-Sys}
\end{table}

\section{Summary}
An amplitude analysis of the decay $D^0 \to K^0_S\pi^0\pi^0$ has been performed  using 20.3 $\rm{fb}^{-1}$ of $e^+e^-$ collision data collected with the BESIII detector at the center-of-mass energy of 3.773 GeV. The BF of $D^0 \to K^0_S\pi^0\pi^0$ is determined to be \BF, using the detection efficiency derived from the amplitude analysis results. The result is consistent with the CLEO result $(1.059 \pm 0.038_{\rm stat.} \pm 0.061_{\rm syst.})\%$~\cite{kspi0pi0}, but the precision is improved by a factor of 5.8.
Combining the FFs listed in Table~\ref{tab:fitresult}, the BFs for the intermediate processes are calculated using $\mathcal{B}_i = {\rm FF}_i \times \mathcal{B}(D^0\to K_S^0\pi^0\pi^0)$. The obtained results are listed in Table~\ref{tab:bfamp}. Significant discrepancies are observed between the results of this work and those reported by CLEO, mainly due to differences in both the amplitude model components and the propagator formalism. In the $M^2(K_S^0\pi^0)$ spectrum, additional contribution from the $(K_S^0\pi^0)_{S-{\rm wave}}$ component is accounted for in this analysis. For the $S$-wave description of the $M^2(\pi^0\pi^0)$ spectrum,  the isobar model was utilized by CLEO, whereas the K-matrix formalism is adopted in this work.

According to the amplitude analysis, the dominant intermediate process is $D^0 \to \bar{K}^{*}(892)^0\pi^0\to K_S^0\pi^0\pi^0$ with a  BF of $(4.10\pm0.10_{\rm{stat.}}\pm0.07_{\rm{syst.}})\times 10^{-3}$. After applying the isospin symmetry assumption to the decays of $\bar{K}^{*}(892)^0\to K^-\pi^+$ and $\bar{K}^{*}(892)^0\to \bar{K}^0\pi^0$, the absolute BF of $D^0 \to \bar{K}^{*}(892)^0\pi^0$ is determined to be $(2.46\pm0.06_{\rm{stat.}}\pm0.04_{\rm{syst.}})\%$. 
Compared with the results listed in Table~\ref{tab:theory}, the result is significantly lower than the predicted results in Refs.~\cite{pole,FAT,TDA} and the CLEO measurement from  the decay $D^0\to K_{S}^{0}\pi^0\pi^0$~\cite{kspi0pi0} by about $3\sigma$. However, it aligns with the value $(2.74\pm0.23_{\rm{stat.}}\pm0.41_{\rm{syst.}})\%$ obtained from the decay $D^0\to K^{-}\pi^+\pi^0$~\cite{kpipi0} but with a precision improved by a factor of 6.0.


Based on the isospin symmetry framework, we measure the ratios to be $\frac{\mathcal{B}((\pi\pi)_{S\text{-wave}} \to \pi^+\pi^-)}{\mathcal{B}((\pi\pi)_{S\text{-wave}} \to \pi^0\pi^0)} = 2.2 \pm 0.6$ and $\frac{\mathcal{B}(f_{2}(1270) \to \pi^+\pi^-)}{\mathcal{B}(f_{2}(1270) \to \pi^0\pi^0)} = 0.7 \pm 0.8$. The $S\text{-wave}$ ratio is in excellent agreement with the predicted value of $2$. The $f_{2}(1270)$ ratio, while central value differs, is consistent with the prediction within $2\sigma$. Our measurements of $\mathcal{B}(D^0 \to K_S^0 (\pi\pi)_{S\text{-wave}})$ and $\mathcal{B}(D^0 \to K_S^0 f_{2}(1270))$ are thus validated by their agreement with the $D^0 \to K_S^0 \pi^+\pi^-$ results from Ref.~\cite{KPsnew2}, supporting the application of isospin symmetry in this analysis.

\begin{table}[t]
\setlength{\abovecaptionskip}{0.cm}
\setlength{\belowcaptionskip}{-0.2cm}
  \begin{center}
    \begin{tabular}{|lcc|}
      \hline
        &\multicolumn{2}{c|}{BF($\times 10^{-3}$)}\\
       \hline
      Intermediate process &This work  &CLEO~\cite{kspi0pi0}\\
      \hline
      $D^0\to \bar K^{*}(892)^0\pi^0, \bar K^{*}(892)^0\to K^0_S\pi^0$ &4.10 $\pm$ 0.10 $\pm$ 0.07 &6.94 $\pm$ 0.61 $\pm$ 0.55\\        
 \hline
       $D^0\to \bar K^{*}(1410)^0\pi^0, \bar K^{*}(1410)^0\to K^0_S\pi^0$ &0.05 $\pm$ 0.03 $\pm$ 0.04 &-\\
 \hline
      $D^0\to \bar K_2^{*}(1430)^0\pi^0, \bar K_{2}^{*}(1430)^0\to K^0_S\pi^0$ &0.16 $\pm$ 0.03 $\pm$ 0.02 &0.05 $\pm$ 0.04 $\pm$ 0.02\\
 \hline
      $D^0\to (K_S^0\pi^0)_{S\rm-wave}\pi^0$ &3.12 $\pm$ 0.55 $\pm$ 0.23 &-\\
 \hline
      $D^0\to \bar K^{*}(1680)^0\pi^0, \bar K^{*}(1680)^0\to K^0_S\pi^0$ &0.38 $\pm$ 0.15 $\pm$ 0.26 &1.18 $\pm$ 0.29 $\pm$ 0.28\\
       \hline
      $D^0\to K_S^0(\pi^0\pi^0)_{S-{\rm wave}}$ &1.50 $\pm$ 0.18 $\pm$ 0.06 &3.06 $\pm$ 0.67 $\pm$ 0.39\\
\hline
      $D^0\to K_S^0f_2(1270), f_2(1270)\to \pi^0\pi^0$ &0.26 $\pm$ 0.06 $\pm$ 0.05 &0.26 $\pm$ 0.10 $\pm$ 0.08\\
 \hline
    \end{tabular}
  \end{center}
      \caption{The comparison of the obtained BFs for intermediate processes with the final state $D^0 \to K_S^0\pi^0\pi^0$ of this work and CLEO. The uncertainties are statistical and systematical, respectively. }
\label{tab:bfamp}
\end{table}

\acknowledgments
The BESIII Collaboration thanks the staff of BEPCII (https://cstr.cn/31109.02.BEPC) and the IHEP computing center for their strong support. This work is supported in part by National Key R\&D Program of China under Contracts Nos. 2023YFA1606000, 2023YFA1606704; National Natural Science Foundation of China (NSFC) under Contracts Nos. 11635010, 11935015, 11935016, 11935018, 12025502, 12035009, 12035013, 12061131003, 12192260, 12192261, 12192262, 12192263, 12192264, 12192265, 12221005, 12225509, 12235017, 12361141819; the Chinese Academy of Sciences (CAS) Large-Scale Scientific Facility Program; CAS under Contract No. YSBR-101; Joint Large-Scale Scientific Facility Fund of the NSFC and the Chinese Academy of Sciences under Contract No.~U2032104; the Excellent Youth Foundation of Henan Scientific Commitee under Contract No.~242300421044; 100 Talents Program of CAS; The Institute of Nuclear and Particle Physics (INPAC) and Shanghai Key Laboratory for Particle Physics and Cosmology; ERC under Contract No. 758462; German Research Foundation DFG under Contract No. FOR5327; Istituto Nazionale di Fisica Nucleare, Italy; Knut and Alice Wallenberg Foundation under Contracts Nos. 2021.0174, 2021.0299; Ministry of Development of Turkey under Contract No. DPT2006K-120470; National Research Foundation of Korea under Contract No. NRF-2022R1A2C1092335; National Science and Technology fund of Mongolia; Polish National Science Centre under Contract No. 2024/53/B/ST2/00975; STFC (United Kingdom); Swedish Research Council under Contract No. 2019.04595; U. S. Department of Energy under Contract No. DE-FG02-05ER41374

\bibliographystyle{JHEP}
\bibliography{references}

\clearpage
\appendix
\include{Appendix}
\include{authorlist_2025-05-14}

\end{document}

%% file: Appendix.tex
\section{\texorpdfstring{$M_{\mathrm{BC}}^{\mathrm{sig}}$ versus $M_{\mathrm{BC}}^{\mathrm{tag}}$ two-dimensional fit}{MBC-sig versus MBC-tag two-dimensional fit}}

\label{2dfit}
The signal yields of DT candidates are determined by a 2D maximum likelihood bin fit to the distribution of $M_{\rm{BC}}^{\rm{sig}}$ versus $M_{\rm{BC}}^{\rm{tag}}$. Signal events with both tag and signal sides reconstructed correctly concentrate around $M_{\rm{BC}}^{\rm{sig}} = M_{\rm{BC}}^{\rm{tag}} = M_{D^0}$, where $M_D$ is the known $D^0$ mass~\cite{PDG}.  We define four kinds of background contributions. Candidates with correctly reconstructed $D^0$(or $\bar D^0$) and incorrectly reconstructed $\bar D^0$(or $D^0$) are BKGI, which appear around the bands $M_{\rm BC}^{\rm sig}$ or $M_{\rm BC}^{\rm tag} = M_{D^0}$. Other candidates appeared around the diagonal are mainly from the $D \bar D$ mispartition and the $e^+e^-\to q\bar{q}$ processes (BKGII). The rest flat background contributions mainly come from candidates reconstructed incorrectly on both sides (BKGIII). The peaking backgrounds come from these events which have the similar daughter particles with our signal mode (BKGIV). The PDFs for the different components used in the fit are given below:
\begin{itemize}
\item Signal: $s(x, y)$,
\item BKGI: $b_1(x,y)$,
\item BKGII:  $b_2(x,y)$,
\item BKGIII: ${\mathcal A}$rgus($x; m_0, c, p$) $\cdot$ ${\mathcal A}$rgus($y; m_0, c, p$).
\item BKGIV:  $p(x,y)$,
 \end{itemize}

The signal shape $s(x, y)$ is described by the 2D MC-simulated shape convolved with a 2D Gaussian. The parameters of the Gaussian function are obtained by a 1D fit on $M_{\rm{BC}}$ in signal and tag sides respectively, and are fixed in the 2D fit. For BKGI and BKGII, $b_{1}(x,y)$ and $b_{2}(x,y)$ are both described by an MC-simulated shape. For BKGIII, it is constructed by an ARGUS function~\cite{ARGUS} in $M_{\rm{BC}}^{\rm{sig}}$ multiplied by an ARGUS function in $M_{\rm{BC}}^{\rm{tag}}$. In the fit, the parameters $m_0$ and $p$ for the ARGUS function~\cite{ARGUS} are fixed at 1.8865 GeV$/c^2$ and 0.5, respectively. For BKGIV, the shape is taken from the inclusive MC sample and we add a Gaussian constraint on its yield in the fit.

 \renewcommand\thesection{\Alph{section}}
\section{The interference between processes}
\label{app:interference}
Table~\ref{tab:inter_roman} shows the Roman numerals for different amplitudes in the nominal model. The interference fit fractions between the amplitudes are listed in Table~\ref{tab:inter}.
The interference between amplitudes calculated by Eq.~(\ref{interferenceFF-Definition}).

\begin{table}[htbp]
    \centering
    \begin{tabular}{|c|c|}
    \hline
           &Amplitude\\
    \hline
       I   & $D^0\to \bar K^{*}(892)^0\pi^0$ \\
       II  & $D^0\to \bar K^{*}(1410)^0\pi^0$\\
       III & $D^0\to \bar K_2^{*}(1430)^0\pi^0$\\
       IV  & $D^0\to (K_S^0\pi^0)_{ S\rm-wave}\pi^0$\\
       V   & $D^0\to \bar K^{*}(1680)^0\pi^0$\\
       VI  & $D^0\to K_S^0(\pi^0\pi^0)_{S-{\rm wave}}$\\
       VII & $D^0\to K_S^0f_2(1270)$\\
    \hline
    \end{tabular}
    \caption{Roman numerals for different amplitudes in the nominal model.}
    \label{tab:inter_roman}
\end{table}

\begin{table}[htbp]
  \centering
  \begin{tabular}{|c|c c c c c c|}
    \hline
    &II &III &IV &V &VI &VII\\
    \hline
    I    &-1.5$\,\pm\,$0.6 &~0.4$\,\pm\,$0.2 &~4.0$\,\pm\,$0.7 &11.1$\,\pm\,$1.7 &~~~9.9$\,\pm\,$1.3 &-2.9$\,\pm\,$0.4\\
    II   &  &-0.2$\,\pm\,$0.1 &-0.6$\,\pm\,$0.2 &~-2.3$\,\pm\,$1.0 &~~-1.1$\,\pm\,$0.5 &-0.4$\,\pm\,$0.2 \\
    III  & &  &-0.6$\,\pm\,$0.1 &~~0.4$\,\pm\,$0.1 &~~-0.8$\,\pm\,$0.3 &~0.4$\,\pm\,$0.2 \\
    IV   & & &  &~-0.4$\,\pm\,$0.6 &~-13.4$\,\pm\,$0.2 &-3.3$\,\pm\,$0.7 \\
    V    & & & &  &~~~7.9$\,\pm\,$0.8 &~0.1$\,\pm\,$0.4 \\
    VI   & & & & &  &-0.0$\,\pm\,$0.0\\
    \hline
  \end{tabular}
  \caption{Interference of each amplitude, scaled by $10^{-2}$ to total amplitude. The uncertainties are statistical only.}
  \label{tab:inter}
\end{table}

%% file: authorlist_2025-05-14.tex
\newcommand{\BESIIIorcid}[1]{\href{https://orcid.org/#1}{\hspace*{0.1em}\raisebox{-0.45ex}{\includegraphics[width=1em]{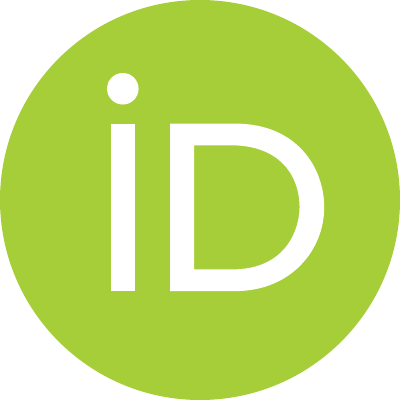}}}}
M.~Ablikim$^{1}$\BESIIIorcid{0000-0002-3935-619X},
M.~N.~Achasov$^{4,b}$\BESIIIorcid{0000-0002-9400-8622},
P.~Adlarson$^{77}$\BESIIIorcid{0000-0001-6280-3851},
X.~C.~Ai$^{82}$\BESIIIorcid{0000-0003-3856-2415},
R.~Aliberti$^{36}$\BESIIIorcid{0000-0003-3500-4012},
A.~Amoroso$^{76A,76C}$\BESIIIorcid{0000-0002-3095-8610},
Q.~An$^{73,59,\dagger}$,
Y.~Bai$^{58}$\BESIIIorcid{0000-0001-6593-5665},
O.~Bakina$^{37}$\BESIIIorcid{0009-0005-0719-7461},
Y.~Ban$^{47,g}$\BESIIIorcid{0000-0002-1912-0374}
% ...
\\
\vspace{0.2cm}
(BESIII Collaboration)\\
\vspace{0.2cm} {\it
$^{1}$ Institute of High Energy Physics, Beijing 100049, People's Republic of China\\
% ...
$^{4}$ Budker Institute of Nuclear Physics SB RAS (BINP), Novosibirsk 630090, Russia\\
% ...
$^{36}$ Johannes Gutenberg University of Mainz, Johann-Joachim-Becher-Weg 45, D-55099 Mainz, Germany\\
$^{37}$ Joint Institute for Nuclear Research, 141980 Dubna, Moscow region, Russia\\
% ...
$^{47}$ Peking University, Beijing 100871, People's Republic of China\\
% ...
$^{58}$ Southeast University, Nanjing 211100, People's Republic of China\\
$^{59}$ State Key Laboratory of Particle Detection and Electronics, Beijing 100049, Hefei 230026, People's Republic of China\\
% ...
$^{73}$ University of Science and Technology of China, Hefei 230026, People's Republic of China\\
% ...
$^{76}$ University of Turin and INFN, (A)University of Turin, I-10125, Turin, Italy; (B)University of Eastern Piedmont, I-15121, Alessandria, Italy; (C)INFN, I-10125, Turin, Italy\\
$^{77}$ Uppsala University, Box 516, SE-75120 Uppsala, Sweden\\
% ...
$^{82}$ Zhengzhou University, Zhengzhou 450001, People's Republic of China\\

\vspace{0.2cm}
$^{\dagger}$ Deceased\\
% ...
$^{b}$ Also at the Novosibirsk State University, Novosibirsk, 630090, Russia\\
% ...
$^{g}$ Also at State Key Laboratory of Nuclear Physics and Technology, Peking University, Beijing 100871, People's Republic of China\\
% ...

}